\documentclass[useAMS,usenatbib]{mn2e}
\usepackage{epsfig,graphicx,latexsym,amsmath,amssymb}
\usepackage[export]{adjustbox}

\usepackage{natbib}
\usepackage{hyperref}
\usepackage{mathrsfs}
\usepackage{lastpage}

\usepackage{color}

\bibpunct[,]{(}{)}{;}{a}{}{,}

\title[Underluminous tidal disruptions]
      {Underluminous tidal disruption events}

\author[P. Amaro Seoane] 
{Pau Amaro Seoane$^{1}$
                        \thanks{E-mail: amaro@riseup.net (PAS)}
\\
Universitat Politècnica de València, Spain\\
Max-Planck-Institute for Extraterrestrial Physics, Garching, Germany\\
Higgs Centre for Theoretical Physics, Edinburgh, UK
}

\def\apj{Astrophys. J.}

\begin{document}
\label{firstpage}

\date{draft \today}

\pagerange{\pageref{firstpage}--\pageref{lastpage}} \pubyear{}

\maketitle

\begin{abstract}
We have evidence of X-ray flares in several galaxies consistent with a a star
being tidally disrupted by a supermassive black hole (MBH).  If the star starts
on a nearly parabolic orbit relative to the MBH, one can derive that the
fallback rate follows a $t^{-5/3}$ decay.
Depending on the penetration factor, $\beta$, a star will be torn apart differently,
and relativistic effects play a role.
We have modified the standard version of the smoothed-particle hydrodynamics
(SPH) code {\sc Gadget} to include a relativistic treatment of the
gravitational forces between the gas particles of a main-sequence (MS) star and
a MBH. We include non-spinning post-Newtonian corrections to
incorpore the periapsis shift and the spin-orbit coupling up to next-to-lowest
order. 
We find that tidal disruptions around MBHs in the relativistic cases are
underluminous for values starting at $\beta \gtrapprox 2.25$; i.e. the fallback
curves produced in the relativistic cases are progressively lower compared to
the Newtonian simulations as the penetration parameter increases. While the Newtonian 
cases display a total disruption, we find
that all relativistic counterparts feature a survival core for penetration
factors going to values as high as $12.05$.
We perform a additional dynamical numerical study which shows that the geodesics of the elements in the star
converge at periapsis. We confirm these findings
with an analytical study of the geodesic separation equation.
The luminosity of TDEs must be lower than predicted theoretically due to the fact that
the star will partially survive when relativistic effects are taken
into account. A survival core should consistently emerge from any TDE with
$\beta \gtrapprox 2.25$. 
\end{abstract}


\section{Motivation}
\label{sec.motivation}

A star passing very close to a massive black hole (MBH) may be torn apart
because of the tidal effects, and the interaction of the stellar debris in the
vicinity of the black hole will give rise to a burst of electromagnetic
radiation. The characteristics of this tidal disruption event (TDE), such as
its temperature, peak luminosity, and decay timescale, are functions of the
mass and spin of the central MBH.  The subsequent accretion of the debris gas
by the black hole produces additional emission, and lead to phases of bright
accretion that may reveal the presence of a MBH in an otherwise quiescent
galaxy \citep[see
e.g.][]{Wheeler1971,Hills75,FR76,CarterLuminet1982,CarterLuminet1983,Rees88,MCD91,MT99,SU99,FB02b,GezariEtal03,WM04},
with rates which vary depending on various factors, but of the order of
$10^{-5}-10^{-6}\,\textrm{yr}^{-1}$ (see e.g. \citealt{Rees1988} and
\citealt{StoneEtAl2020} for a recent review on the rates and characteristics).
These phenomena can be used as a probe of accretion physics close to the event
horizon \citep{Brenneman2013,Reynolds2014}.

Many disruption candidates have already been detected with ROSAT, Chandra,
Swift (see e.g. \href{https://tde.space}{https://tde.space}), and the ZTF
\citep{vanVelzenEtAl2020} and the number will surge with upcoming transient
surveys like the Large Synoptic Survey Telescope (LSST), SRG/eROSITA
\citep{KhabibullinEtAl2014}, as well as the ESA L2 mission Athena
\citep{Athena+Whitepaper}. 

A conondrum related to optically-discovered TDEs is that their color
temperatures are significantly below the values predicted by theoretical models
\citep{GezariEtAl2012,ChornockEtAl2014,HoloienEtAl2014,vanVelzenEtAl2014,ArcaviEtAl2014}.
Observations depict a temperature and bolometric luminosity well below
theoretical predictions based on accretion, and based on the same model, the
derived black-body emission radius implies an orbital motion below the expected
theoretical values. In general, the fallback model requires masses much less
than a solar mass in order to explain the difference in luminosity of the
observed flares and the theoretical expectations.

Different theoretical models have been put forward to explain this fact.  The
work of \cite{LiEtAl2002} suggested that the low luminosity may indicate that
the disrupted star is a brown dwarf or a planet.  An alternative explanation is
that the assumption that the gas immediately circularizes when it comes back
close the MBH is not accurate, and could trigger internal shocks that would
result into a reduced luminosity. In particular, \cite{PiranEtAl2015} suggested
that the released energy via internal shocks is responsible for the observed optical TDE candidates.
More recently, \cite{ZhouEtAl2020} argued that the disk does not circularize
and remains eccentric, which as a consequence leads the orbital energy of the
stellar debris to be advected on to the MBH without being radiated away.

In this work we show with a set of smoothed-particle hydrodynamics (SPH)
simulations with relativistic corrections that unbound stars lead to partial
disruptions, which naturally explain the difference in the observed luminosity,
for penetration values as deep as $\beta=12.05$.

\section{Relativistic implementation}

Relativistic effects have been considered in the related literature by e.g.
\cite{TejedaEtAl2017}, which implemented a relativistic description of the
evolution of the hydrodynamical elements with a quasi-Newtonian treatment of
the fluid's self-gravity.  

Earlier this year, Ryu and collaborators presented a series of works which also
address TDEs in a relativistically fashion. For this approach they depart from
the intrinsically-conservative GR hydrodynamical numerical code of
\cite{NobleEtAl2009} designed to study magnetohydrodynamic (MHD) turbulence in
accretion disks around MBHs. To study TDEs, they assume that spacetime is
Schwarzschild plus contributions from the star self-gravity, and the dynamics
of the star is described by hydrodynamics, solving the general-relativistic
energy-momentum equations in the Schwarzschild background. This hence implies
that in the absence of material forces the fluid elements strictly follow
geodesics. The self-gravity of the star is described in the weak field, taking
into account only the Newtonian gravitational potential.  They then evolve the
hydrodynamical equations in a frame where the metric is nearly flat and  move
the whole system in a rigid way along the orbit using parallel transport of the
local frame. Hence, it is the trajectory of the system what ``sees'' the
Schwarzschild metric but the fluid elements almost live in a flat spacetime.

More precisely, they consider a modified metric, $\tilde{g}_{\mu\,\nu} =
g_{\mu\,\nu} + h_{\mu\,\nu}$, with $g_{\mu\,\nu}$ the Schwarzchild
metric and $h_{\mu\,\nu}$ accounts for the self-gravity of the star. They
assume the self-gravity is weak, so that the only non-zero component of the
self-gravity perturbations is the time-time one: $h_{tt} = -2 \Phi_{sg} /
c^2$, where $\Phi_{sg}$ is the Newtonian potential of the star, in the sense
that it satisfies a Poisson equation where the mass density is replaced by the
star proper rest-mass density. The procedure to incorporate this self-gravity
is a bit more intrincate than adding it to the Schwarzschild metric. The
assumptions made to compute the self-gravity contribution include that the
metric of Schwarzschild should not deviate from the Minkowski metric. Here is
where the intrincacy mentioned comes, since they need a parallel-transported
tetrad adapted to the star as mentioned before, so that in that frame the
assumptions made are valid. It is important to note that they separate the
problem of solving the hydrodynamical equations from the motion of the star
around the MBH. This can be envisaged as having a frame center at the (center
of mass) star where the metric, in an orthonormal basis, is exactly Minkowski
(deviating as we move from the center of mass). Then they solve the
hydrodynamical equations in this frame and the motion of the star is ``rigid''
(only the center of mass moves) according to the parallel-transport equations
for the orthonormal basis \citep[see][]{RyuEtAl2020b}.  In their calculations
all stars have net bound orbits by an amount of the order $\sim 10^{-10}c^2
\sim 10^{-3} (\sigma^2/2)$, where $\sigma$ is the bulge dispersion and $c$ the
speed of light.

With this scheme they investigate TDEs in four additional works. In
\cite{RyuEtAl2020a} they find that the critical pericenter distance for full
disruptions is enhanced by up to a factor of $\sim 3$ as compared to the
Newtonian case, and that it depends on the mass of the star in a non-trivial
way \citep[see][for previous work]{GuillochonRamirez-Ruiz2013}.  The results of
\cite{RyuEtAl2020c} regarding partial disruptions show that due to the little
mass distributed at low energies, late-time fallback is suppressed. The mass
return rate should then be $\propto t^{-p}$ with $p\in [2,\,5]$ in partial
disruptions.  In \cite{RyuEtAl2020d} they show that relativistic effects induce
width  delays in the debris energy so that the magnitude of the peak return
rate decreases.  These results had already been pointed out by the previous
work of \cite{IvanovChernyakova2006,Kesden2012,ServinKesden2017}, although Ryu
and collaborators provided quantitative corrections to these previous
treatments.  In \cite{KrolikEtAl2020} they discuss the event rates and the fate
of the rest of the star which is not disrupted (i.e. the amount of mass still
inside the computational box when they stopped the simulation), which might
interact with the MBH on a second periapsis passage or rejoin the stellar
cluster. As we will see, in this work we find a surviving core in all relativistic
simulations, which is in contradiction with a total disruption.

In this work we modify the acceleration computation of {\sc Gadget}
\citep{Springel2005} to include relativistic corrections, which are based on
the post-Newtonian (PN) formalism for the interaction between two bodies (in
our case each of the hydrodynamical particles and the MBH).  This means that we
simply add relativistic correcting terms to the Newtonian gravitational forces
calculated between the MBH and the hydrodynamical particles that form the star
during the whole simulation, which initially is set on a completly unbound
orbit.  This approximation allows us to capture the relativistic effects while
allowing us to study the evolution of the star to larger radii without any
other approximation than those inherent to SPH methods and the post-Newtonian
expansion, valid in this regime of low (but yet relativistic) velocities. In
this regard, our scheme is self-consistent and all phenomena related to
relativistic effects and hydrodynamics emerge naturally by integrating the
system.

The relative acceleration, in the center-of-mass form, i.e. following the scheme of \cite{BremAmaro-SeoaneSpurzem2013}, including all PN
corrections can be written in the following way:

\begin{equation}
 \frac{d \vec v}{d t} = - \frac{G m}{r^2}[(1+A)\,\vec n + B \vec v \,] + \vec C_{\rm 1.5,SO}
\label{eq.acc}
\end{equation}

\noindent
In this equation $\vec v = \vec v_1 - \vec v_2$ is the relative velocity
vector, $m = M_\bullet + m_{\rm gas}$ the total mass, with $M_\bullet$ the mass
of the MBH and $m_{\rm gas}$ the mass of a gas particle, $r$ the separation and
$\vec n = \vec r/r$. $A$ and $B$ are coefficients that can be found in
\cite{BlanchetIyer03}. Since we are modelling extremely light gas particles
around a MBH, we adopt the terms in the limit $m_{\rm gas} = 0$.  We consider
the leading order spin-orbit interaction, with the term $\vec C_{\rm 1.5,SO}$
in which the subscript SO stands for spin-orbit coupling, which can be found in
\citep{TagoshiEtAl01,FayeEtAl2006}, and the first post-Newtonian correction to
periapsis shift. We do not include dissipative terms because, contrary to an
extreme-mass ratio inspiral \citep{Amaro-SeoaneLRR2012}, the star only has one periapsis passage,
and the gravitational radiation can be neglected. All the PN interactions are only considered between a gas
particle and the MBH and are evaluated at all times during the whole integration. In all simulations with
spin we set the dimensionless spin vector to $\vec a = (0.7,\,0.7,\,0)$, so we
get maximum precession of the orbit of the star orbit lying in the X--Y plane.

The implementation of these relativistic terms follows the prescription given
in the work of \cite{KupiEtAl06}, which was the first work published about the
inclusion of post-Newtonian corrections in the context of stellar dynamics.
The addition of the spin to the problem was presented, also for the first time,
in a stellar-dynamics context in \cite{BremAmaro-SeoaneSpurzem2014}.  Both, the
periapsis shift and the spin terms have been tested in detail, and partially
published in the work of \cite{BremAmaro-SeoaneSpurzem2014} with a series of
comparisons with the semi-Keplerian approximation of \cite{Peters64}.

\section{Initial conditions}

In all simulations the mass of the MBH is $m_{\bullet}=10^6\,M_{\odot}$, the
mass of the star is $m_{\star}=1\,M_{\odot}$ and it is set on an unbound,
parabolic trajectory around the MBH, placed at the focus. It must be noted that
while bound orbits are less ``expensive'' computationally, the most natural
orbits are unbound ones, i.e. parabolic or hyperbolic, because we do not expect
the region of phase-space close to the MBH to produce bound orbits, at least in
a Milky Way-like galaxy\citep[see
e.g.][]{Amaro-SeoaneLRR2012,BaumgardtEtAl2018,SchoedelEtAl2018,Gallego-CanoEtAl2018}.

When the stars approach the MBH it will experience strong tidal forces whenever
the work exerted over the star by the tidal force exceeds its own binding
energy, (all energies are per unit mass), which is 

\begin{equation}
E_{\rm bind}=\frac{3\,G\,m_{\star}}{(5-n)r_{\star}},
\label{eq.Ebind}
\end{equation}

\noindent
where $n$ is the polytropic index \citep{Chandra42}, $m_{\star}$ the mass of
the star. This allows us to introduce a typical radius for this to happen, the
tidal radius $r_t$. Considering $r_{\star}\ll r_{\rm t}$,

\begin{equation}
r_t=\Bigg[\frac{(5-n)}{3}
\frac{{m}_{\bullet}}{m_{\star}}\Bigg]^{1/3} 2\,r_{\star}.
\label{eq.r_tid_bind}
\end{equation}

\noindent
With ${m}_{\bullet}$ the mass of the MBH. For a solar-type star, considering an $n=3$ 
polytrope, and ${m}_{\bullet}=10^6\,M_{\odot}$, we have that

\begin{equation}
r_{\rm t} = 110\,R_{\odot}\sim 0.51\,\textrm{AU}.
\label{eq.r_tib_bind_b}
\end{equation}

The initial distance of the star to the MBH is set to 20 times the axis of
symmetry of the parabola, i.e. the pericentric distance between the MBH and the
star. In order to investigate the fate of the bound material to
the star and the fallback rate, we have chosen a series of trajectories with
different penetration factors $\beta$, which is defined to be the ratio between the
tidal radius and the distance of periapsis, $1.5,\,2,\,3,\,4,\,5$ and $9$, and
run for each case (i) a Newtonian simulation, (ii) a relativistic simulation
taking into account only the periapsis shift of the SPH particles and (iii) a
relativistic simulation taking into account this effect and the spin
correction. 

\begin{figure*}
\resizebox{\hsize}{!}
          {\includegraphics[scale=1,clip]{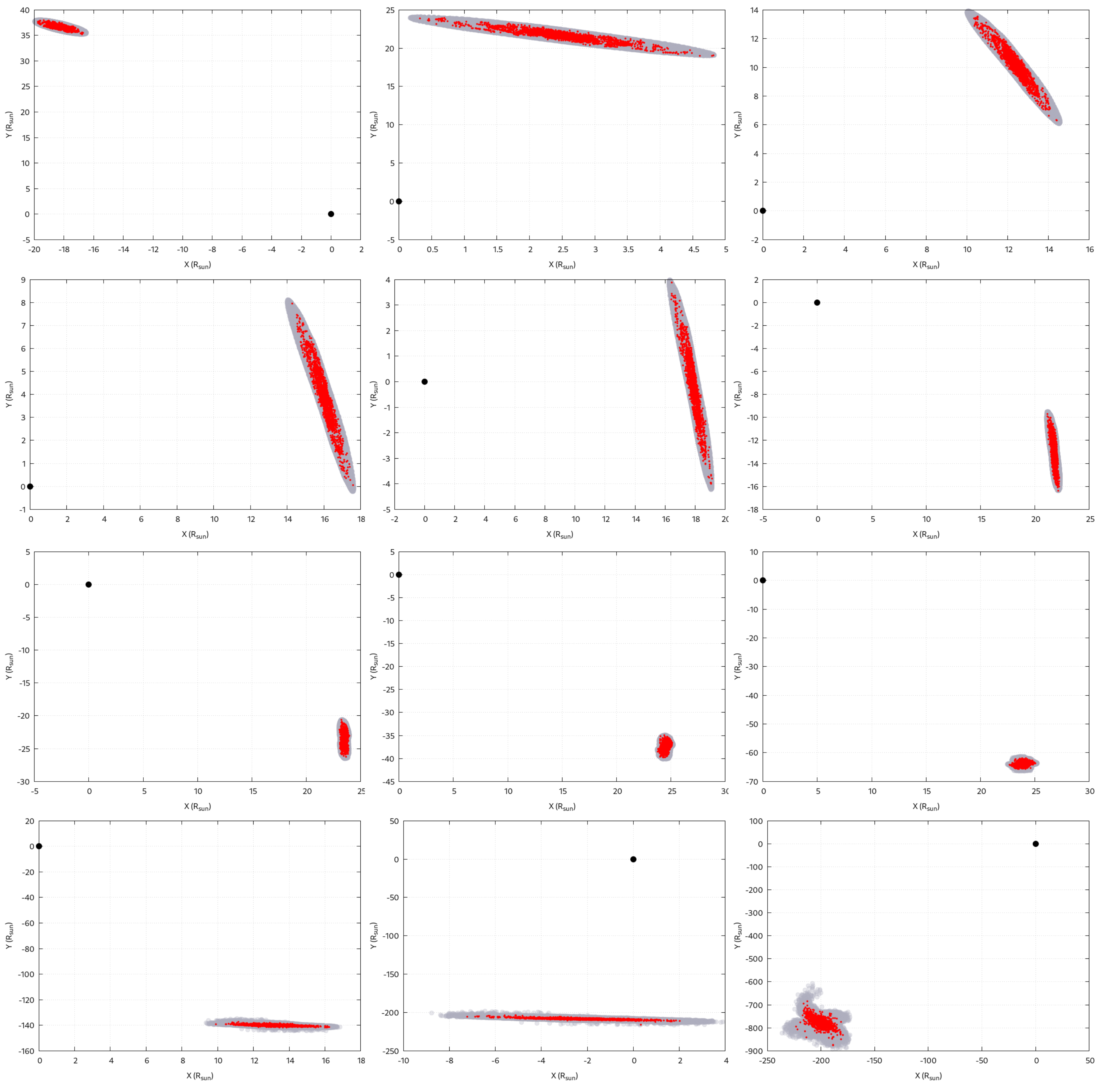}}
\caption
   {
Tidal disruption of the $\beta=12.05$ case with relativistic
corrections. We can observe, from the top to the bottom,
right to left, how the star becomes more elongated along
its parabolic orbit around the MBH, represented with a
black dot at the origin. After the stretching and compression
of periapsis, the star recovers a somewhat spherical architecture to again be stretched. At much longer timesteps,
which corresponds to the last panel, the star is again more
spherical although with deformations. We do not show all gas
particles that we used in the simulation for clearity, and
in red the densest regions in the star.
   }
\label{fig.Disruption}
\end{figure*}

It must be noted that the value for the penetration factors has been estimated
by assuming a point-particle trajectory.  However, in the relativistic cases,
an initially assigned value for $\beta$ diverges as the star progresses in its
orbit towards the MBH as a function of the penetration factor. When the
extended star achieves the vertex of the parabola, the penetration factor has
differed from the initially chosen value. Hence, we initially set the star in
that point-particle trajectory for those specific $\beta$ values, and we derive
the real penetration factor when it reaches the verteces of the relativistic
cases.  These are $\beta=1.64,\,2.26,\,3.62,\,5.15,\,6.83$ and $12.05$, and we
use them in the Newtonian cases as well so as to be able to compare the
results. These values in turn correspond to the following distances of
periapsis: $r_{\rm peri} \sim
0.32,\,0.22,\,0.14,\,0.1,\,0.07,\,0.04~\textrm{AU}$. For a parabolic orbit, the
velocity of the star at periapsis is $v_{\rm peri}=\sqrt{2\,\mu/r_{\rm peri}}$,
with $\mu=G\,m_{\bullet}\simeq 1.3\times 10^{26} m^3\,s^{-2}$, so that the
corresponding velocities are $v_{\rm peri} \sim 7.37\times 10^7,\, 8.88 \times
10^7,\,1.11 \times 10^8,\,1.32\times 10^8,\,1.57\times 10^8,\,2.08\times
10^8\,m/s$ which in units of the speed of light $c$ are, respectively,
$0.24,\,0.29,\,0.37,\,0.44,\,0.52,\,0.69$. In Fig.~(\ref{fig.Disruption}) we show 12 different 
snapshots in the evolution of a representative case with relativistic
corrections, $\beta=12.05$. We have displayed in red the
densest regions of the star.

The stars are modelled as main-sequence (MS) stars with a polytrope of index
$3$ constructed initially following the method of \cite[][in particular, see
their on-line complements]{FB05}. We employ half a million particles to
construct the polytrope, which is enough to solve the tidal disruption process.
Increasing the number of particles by less than an
order of magnitude (or a factor of 10) does not
necessarily lead to a significant improvement of the simulation \citep[see section 3
of][]{Rasio2000}, as we show in section~(\ref{sec.resolution}), in particular of
the stellar density and temperature profile. We adopt a fixed softening length for the gas particles of $0.01\,R_{\odot}$. Taking into account that the deepest penetration factor we have used is of $12.05$, and that the
tidal radius is of $110\,R_{\odot}$, as we can see in Eq.~(\ref{eq.r_tib_bind_b}),
this means that the closes periapsis distance is of $\sim 9.13\,R_{\odot}$; i.e.
almost three orders of magnitude larger than our softening, so that the
simulation is well resolved. The has smoothing is adaptive, whereby the number of neighbours with which each particle interacts is constant, and set to 50. 

\section{Quantitative analysis}

One important aspect in the process is the evolution of the star after the first
periapsis passage.  To determine which part of the gas particles in the
simulation is still bound or unbound to the star, we follow the following
prescription \citep[based in the work of][]{LaiEtAl1993,FulbrightEtAl1995}: So
as to decide whether a gas particle $i$ is bound to the star, we calculate the
specific energy of this particle relative to the star,

\begin{equation}
\epsilon_i = u_i + \frac{1}{2} v^2_{\rm rel} - \sum_j \frac{G m_j}{r_{ij}},
\label{eq.specific}
\end{equation}

\noindent
where $u_i$ and $v_{\rm rel}$  are respectively
the internal energy and the relative velocity of gas particle $i$ to the center
of mass velocity of all the gas particles belonging to the star. The potential
part is summed up over all star particles $j$. If $\epsilon_i > 0$, the
particle is considered unbound from the star, otherwise bound. In the first
step of the iteration, all particles are assumed to be star particles. After
evaluating Eq. \ref{eq.specific}, particles are reassigned to be either still
part of the star or unbound. In the next step, the specific energy is evaluated
with respect to the reduced fraction of star particles. We stop the iteration
when reassignments to the unbound component cannot be made.  After the
iteration is complete we check which part of the gas that is no longer
gravitationally bound to the star is on Keplerian orbits around the MBH or
completely unbound from the system. This part of the gas is then considered for
the luminous fallback onto the MBH.

\begin{figure*}
\resizebox{\hsize}{!}
          {\includegraphics[scale=1,clip]{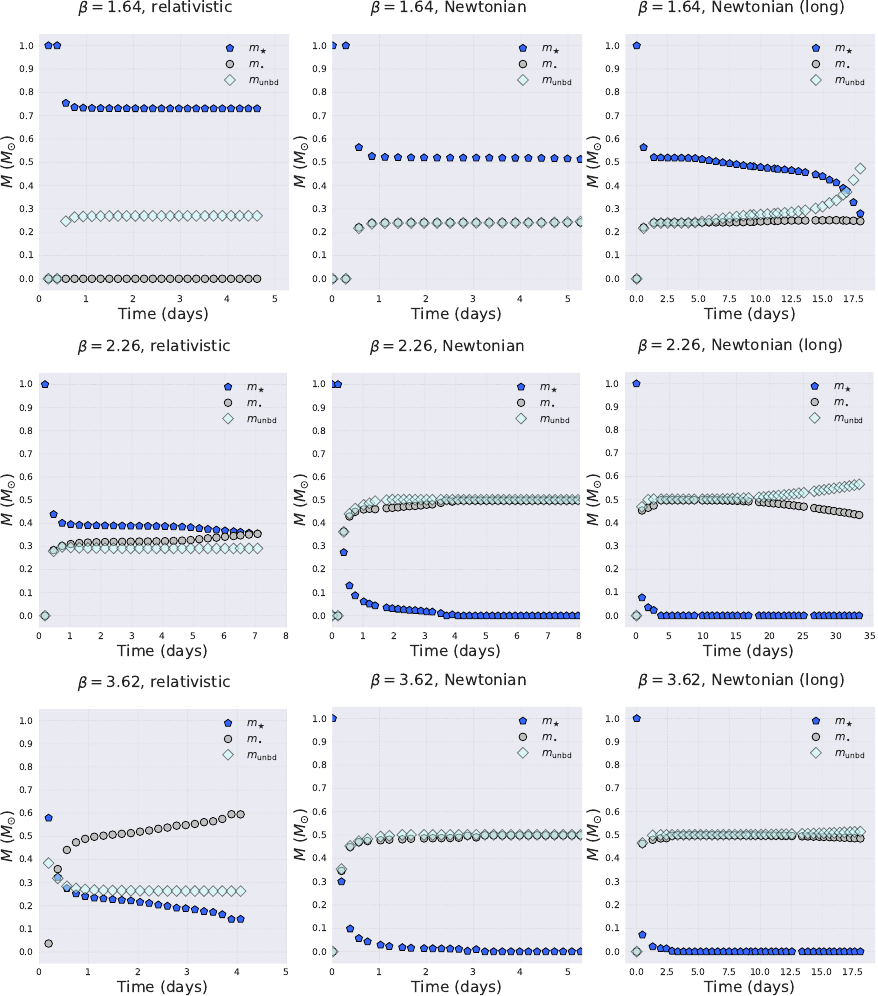}}
\caption
   {
Amount of bound and ejected stellar mass for $\beta=1.64$ as a function of time.
In blue pentagons we depict bound stellar mass to the star, i.e. this represents
the evolution of how much of the star remains after the TDE. In grey circles the
amount bound to the MBH, and in cyan squares the unbound stellar mass, i.e. the
mass of the star which is ejected. The left and middle panels correspond to the
relativistic and Newtonian simulations set to the same limits, so as to be able
to compare. The right panel is the Newtonian case integrated significantly further
than in the left panel (labelled ``long'').
   }
\label{fig.Bound_Ejected_Mass_Beta1p64_to_3p62}
\end{figure*}

In Fig.~(\ref{fig.Bound_Ejected_Mass_Beta1p64_to_3p62}) we show the fate of the
material stripped (bound) from (to) the star for the different penetration
factors mentioned before up to 3.62. The scaling of the first two panels in all
these figures is the same in order to be able to compare better.  In all
figures we also add a third panel showing the long-term evolution of the
Newtonian cases.  We can see from these figures that the amount of bound mass
to the star, i.e.  the survival star, is in all cases larger in the
relativistic simulations than in the Newtonian counterparts.  For the
relativistic cases, about $70\%,\,40\%,\,20\%$ of the star survives the
disruption after one day for the first three values of
$\beta=1.64,\,2.26,\,3.62$, while in the Newtonian simulations, this quantity
is only about $50\%,\,5\%,2\%$ These results are general in agreement with the
amount of mass of the survival core found by the work of \cite{RyuEtAl2020c}.

\begin{figure*}
\resizebox{\hsize}{!}
          {\includegraphics[scale=1,clip]{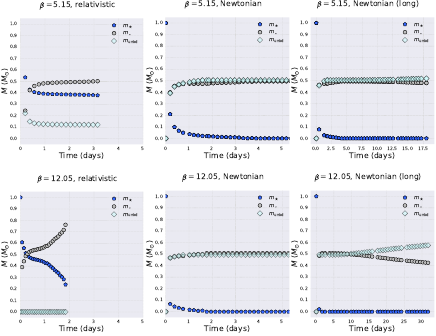}}
\caption
   {
Same as Fig.~(\ref{fig.Bound_Ejected_Mass_Beta1p64_to_3p62}) but for the two more extreme
penetration factors, $\beta=5.15$ and $12.05$.
   }
\label{fig.Bound_Ejected_Mass_Beta5p15_and_12p05}
\end{figure*}

In Figure~(\ref{fig.Bound_Ejected_Mass_Beta5p15_and_12p05}) we depict the same
as in Fig.~(\ref{fig.Bound_Ejected_Mass_Beta1p64_to_3p62}) for the two most
extreme penetration factors, $\beta=5.15,\,12.05$ (this last penetration factor
the star is still at a distance of two Schwarzschild radii from a Schwarzschild
MBH of mass $10^6\,M_{\odot}$). In this case, the amount of bound material to
the star in the Newtonian cases further decreases as compared to larger $\beta$
values. It is of about $1\%$ and $0\%$ respectively. However, the relativistic
cases show a much larger survival stellar object, with a mass of about $40\%$.  In the relativistic simulations the
amount of bound material to the star is larger than in their Newtonian
counterparts, even at the smallest value of $\beta$.

In this first case, for $\beta=1.64$, we observe in the Newtonian case of
Fig.(\ref{fig.Bound_Ejected_Mass_Beta1p64_to_3p62}) that a significant amount
of matter of the star survives the close interaction with the MBH.  Indeed,
\cite{GuillochonRamirez-Ruiz2013} estimated that (Newtonian) TDEs have 100\%
disruption only for penetration factors $\beta > 1.85$.  Recently,
\cite{MilesEtAl2020} have studied (Newtonian) partial disruptions that
corroborate the fallback rate proportion of $\propto t^{-9/4}$. Other
scenarios, like the progressive disruption of a star as result of a tidal
separation, however, lead to different eccentricities which predict a different
scaling of $\propto t^{-1.2}$ \cite{Amaro-SeoaneMillerKennedy2012}. Also,
\cite{RyuEtAl2020c} find $\propto t^{-p}$ with $p\in [2,\,5]$ depending on the mass of the star
and the role of their relativistic implementation.

\begin{figure}
\resizebox{\hsize}{!}
          {\includegraphics[scale=1,clip]{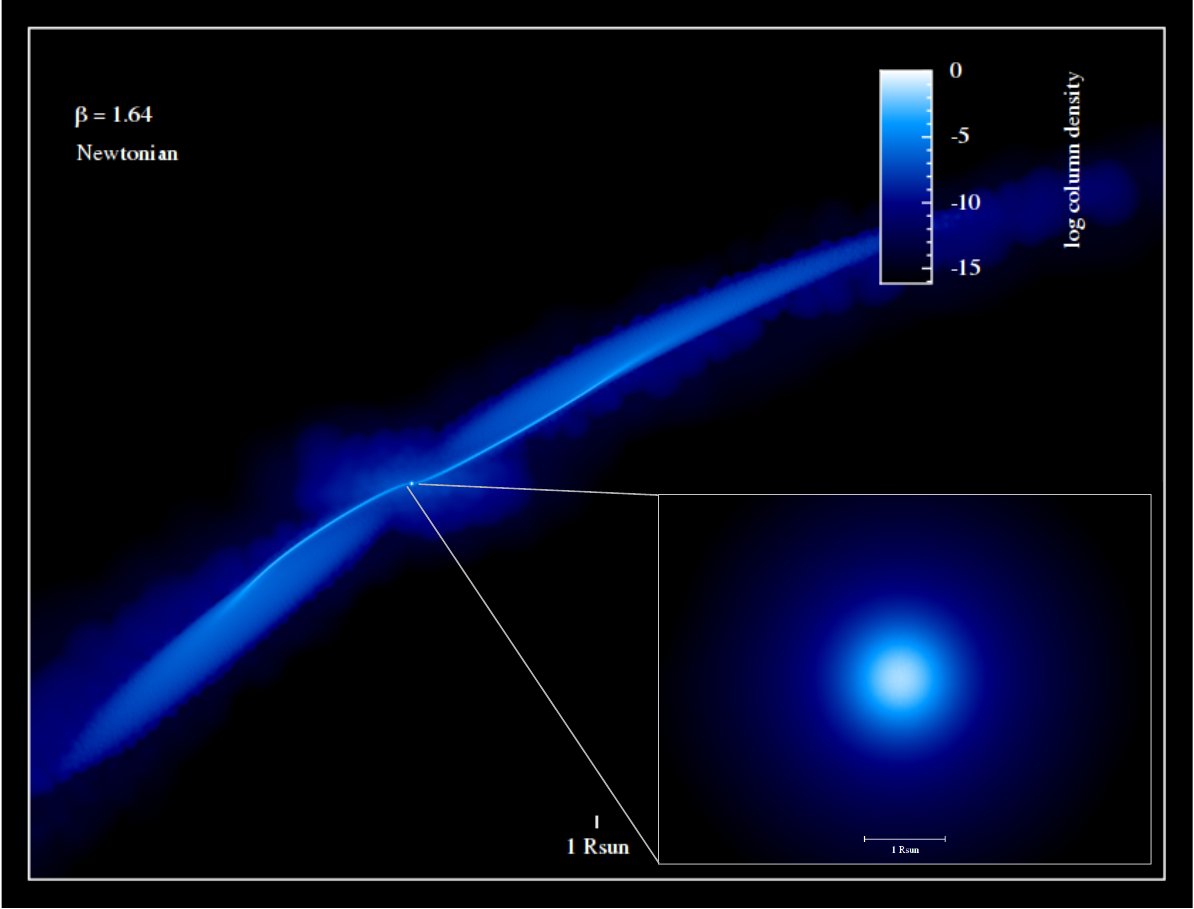}}
\caption
   {
Last snapshot of the Newtonian simulation for $\beta=1.64$. Both panels have a reference
bar showing the length of a solar radius (``Rsun'') to set the scale. The bar depicting the
logarithm of the column density of gas in the zoom has been set to the minimum and
maximum values of $-3.43$ and $0$, respectively for a better identification of the structure.
   }
\label{fig.SurvivalCoreBeta1p64_N}
\end{figure}

In Fig.~(\ref{fig.SurvivalCoreBeta1p64_N}) we show the last snapshot of the
Newtonian simulation with $\beta=1.64$ using the visualization tool of
\cite{Price2007} to render the gas particles. Embedded in the figure we have
added a zoom of the area corresponding to what seems to be the remaining core
of the star.  However, this is a transient feature, as we can see in the
uppermost, right panel of Fig.~(\ref{fig.Bound_Ejected_Mass_Beta1p64_to_3p62}).
We integrated the system for up to some $\sim 18$ days from the starting point,
and we can see that the amount of mass decays very quickly with time, while the
amount of matter of the original star bound to the MBH is kept constant.  This
episodic core will not be bound to the star at later times. However, it appears
later as an enhanced fallback of debris in the first panel of
Fig.~(\ref{fig.FallbackBetas}) at later times, as we will explain in the next section.

\section{Fallback rate}

\begin{figure*}
\resizebox{\hsize}{!}
          {\includegraphics[scale=1,clip]{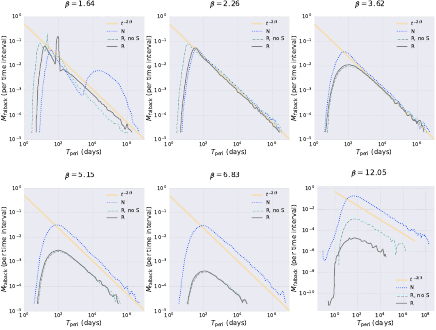}}
\caption
   {
Fallback rate for a polytrope of index $3$, mass $1\,M_{\odot}$, radius $1\,R_{\odot}$ and
$500,000$ particles. The curves delimit the upper part of histograms relative
to the amount of bound stellar mass per time interval, distributed in intervals
of $T_{\rm peri}$, in days. The dotted, blue curve corresponds to the Newtonian (N)
simulations, the long-dashed, grey curves to the relativistic cases without
spin (R, no S) and the solid, grey curves to the relativistic runs taking into
account spin corrections (R). The solid, light-orange line depicts the power-law described in
the work of Rees in which the fallback rate is $\propto t^{-5/3}$. The
reason for the exponent being $-2/3$ and not $-5/3$ is to correct for the
logarithmic representation of the results, since the derivative of the mass
$M$ respect to the logarithm of time, $t$, $dM/d(\log(t)) \propto t \cdot dM/dt$.
Hence, for a relation such as $dM/dt \propto t^{-x}$, $dM/d(\log(t)) \propto
t^{-x+1}$, and so $-5/3+1 = -2/3$.
   }
\label{fig.FallbackBetas}
\end{figure*}

\vspace{0.3cm}

So as to test the implementation of the orbit and the behavior of the SPH star,
we calculate the fallback rate on to the MBH. For this, we calculate the
required time for the bound debris to come back again to periapsis by
estimating the specific energy of each particle, $E = G\left(M +
m\right)/(2\,a)$, with $G$ the gravitational constant, $M$ and $m$ the masses
of the MBH and one SPH particle, respectively.

From the angular momentum, one can derive that

\begin{equation}
e = \sqrt{1 + E \left(\frac{L}{\mu} \right)^2},
\label{eq.}
\end{equation}

\noindent
where we have introduced $\mu = G\,M$ and neglect the contribution of $m$. If we
define $\Delta t$ as the ellapsed time between the first periapsis passage and the
current position of the particle's position, at a radius $R$ from the MBH and time $T$, the
necessary time for the next periapsis passage is $t_{\rm peri}=T - \Delta t$.

This distance is

\begin{equation}
R = a\,\left(1-e\cdot \cos \epsilon\right),
\end{equation}

\noindent
with $\epsilon$
the eccentric anomaly; so that $\epsilon < \pi$ for outbound motion and
$\epsilon > \pi$ for inbound motion.  The mean anomaly can be calculated as

\begin{equation}
M = 2\pi \delta t/T = \epsilon -e \cdot \sin{\epsilon},
\end{equation}

\noindent
and allows us to calculate the ellapsed time since the last periapsis passage
(at $M=0$).  We integrate the star's orbit until its center of mass has
traveled far away from the tidal radius out to $3000 R_\odot$, i.e.  $\sim
12730\,R_{\rm Schw}$, with $R_{\rm Schw}$ the Schwarzschild radius of the MBH.
From this point on we assume the gas particles to be traveling on independent,
non-interacting Keplerian orbits solely determined by their orbital energy and
angular momentum. From these values we compute the classical time until the
subsequent pericenter passage, $t_{\rm peri}$.

We present the results as fallback curves which are mass histograms over
$t_{\rm peri}$ in Fig.~(\ref{fig.FallbackBetas}) for different values of the
penetration factor $\beta$.  We can see that the lower value of $\beta$ leads
to a drop in the Newtonian fallback between $10^2$ and $10^4$ days, which can
be envisaged as a result of matter falling back more quickly. This forms the
depression around the later enhanced fallback. This is matter which is bound to
the MBH, not to the original star, which has started to become bound to itself
after the evolution. The relativistic cases also display this feature but at
much earlier times and with much smaller depressions. This partial disruption
leads to fallback values in the Newtonian case which are similar to the
relativistic ones, as in the next value of $\beta=2.26$. From that value
upwards, the Newtonian cases lead to fallback values significantly higher than
the relativistic ones, starting with about half an order of magnitude up to
about five orders of magnitude for the deepest penetration and the spin case.
We can observe only a clear effect of the spin when we go to extreme
penetration values, the lowermost, right panel, with $\beta=12.05$. 

\begin{figure*}
\resizebox{\hsize}{!}
          {\includegraphics[scale=1,clip]{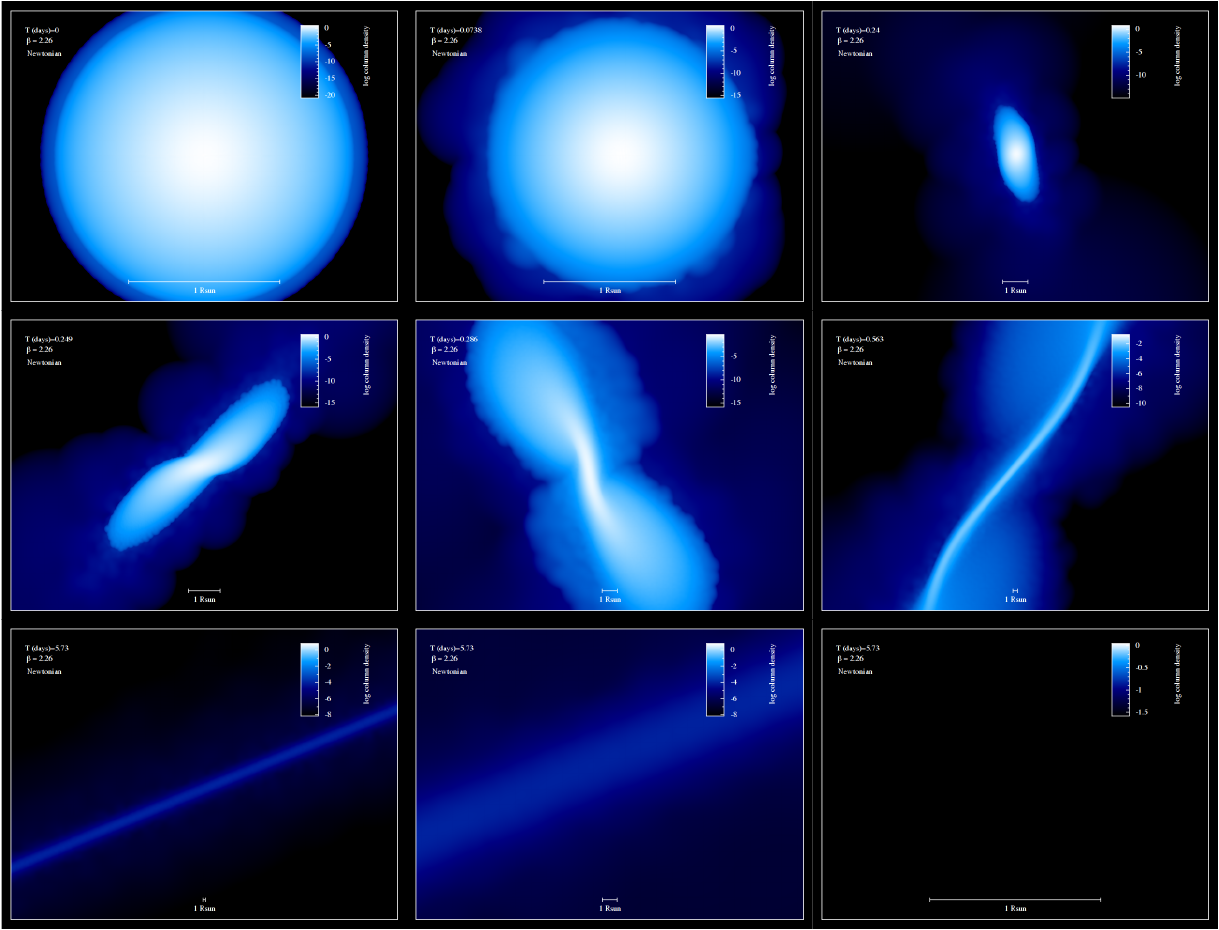}}
\caption
   {
Mosaic of different instants of time corresponding to the Newtonian simulation
of penetration factor $\beta=2.26$. From the top to the bottom, left to right,
We show six different moments in the evolution of the star in its reference
frame, more specifically at (approximately)
$T=0,\,0.07,\,0.24,\,0.25,\,0.29,\,0.56$ and $5.73$ days.  As a reference
point, we show the length corresponding to one solar radius in each panel, and
the (logarithm) of the column density of the star at top on the right.  Both
the zoom factor and the depth of the logarithmic scale in the column density
have been chosen in each frame to show the most interesting features, and they
do not necessarily match from frame to frame. The last three panels depict
the last moment, $5.73$ days, at different zoom levels of the densest region.
The last one shows an empty area because the depth of the column density is
set to values not found.
}
\label{fig.SPH_mosaic_beta_2p26_N}
\end{figure*}

\begin{figure*}
\resizebox{\hsize}{!}
          {\includegraphics[scale=1,clip]{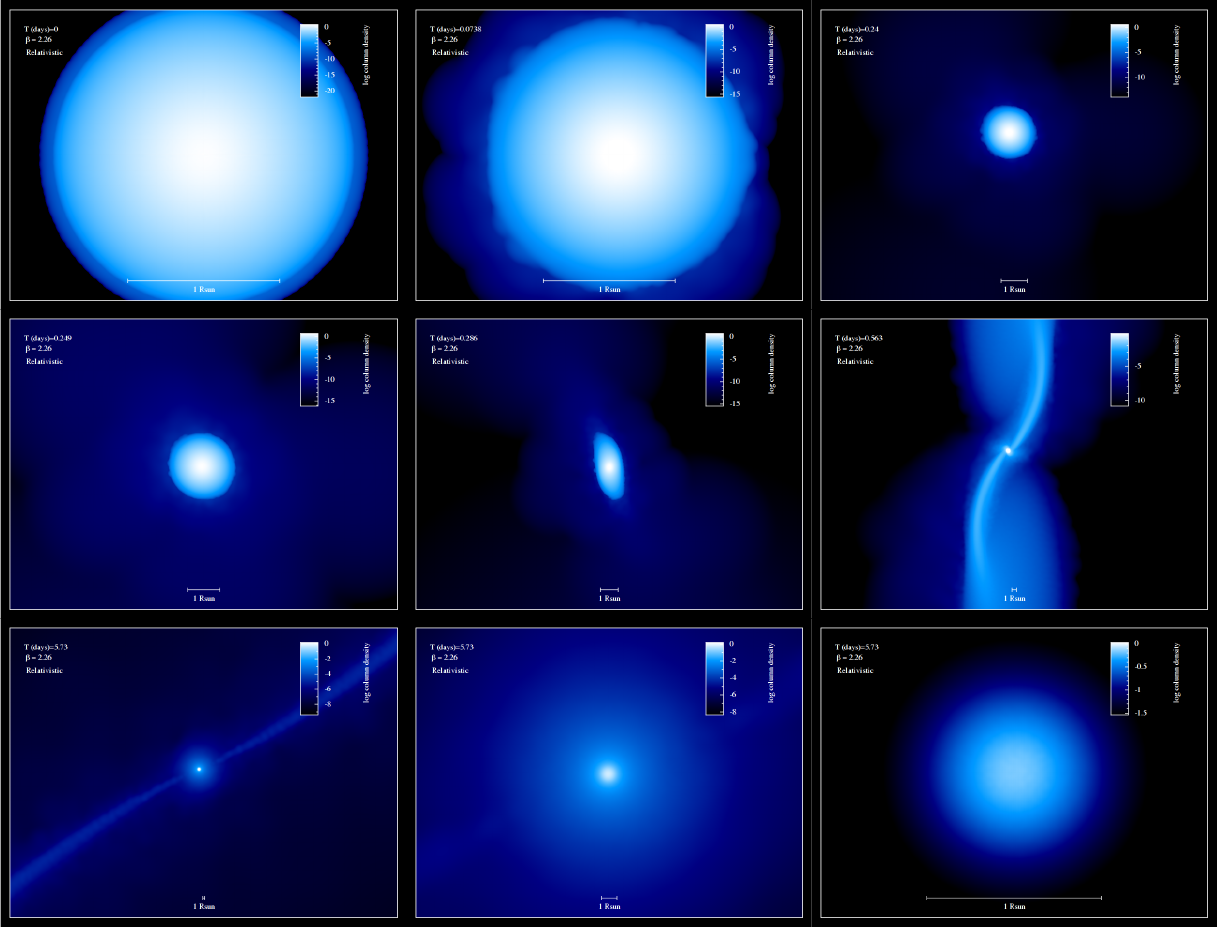}}
\caption
   {
Same as Fig.~(\ref{fig.SPH_mosaic_beta_2p26_N}) for the relativistic
counterpart. Whilst in the previous figure the last panels show a thread-like
distribution of the gas debris, in the relativistic counterpart we see a
survival core with a size of about half of the initial star. Both, the lengths
and the density of the gas have been set to almost identical values to those shown
in Fig.~(\ref{fig.SPH_mosaic_beta_2p26_N}), so as to be able to compare panel by panel
of both figures.
   }
\label{fig.SPH_mosaic_beta_2p26_R}
\end{figure*}

In Fig.~(\ref{fig.SPH_mosaic_beta_2p26_N}) we show a mosaic with nine different
snapshots in the evolution of the Newtonian case of $\beta=2.26$. As we zoom
in, we can see that at later times, 5.73 days, no surviving core is left. This
situation changes completely when we consider the relativistic corrections, as
we can see in its counterpart, Fig.~(\ref{fig.SPH_mosaic_beta_2p26_R}), which
takes into account the post-Newtonian correcting terms. We can clearly see the
survival of a core which is bound to the original star, as shown in
Fig.~(\ref{fig.Bound_Ejected_Mass_Beta1p64_to_3p62}).

\begin{figure*}
\resizebox{\hsize}{!}
          {\includegraphics[scale=1,clip]{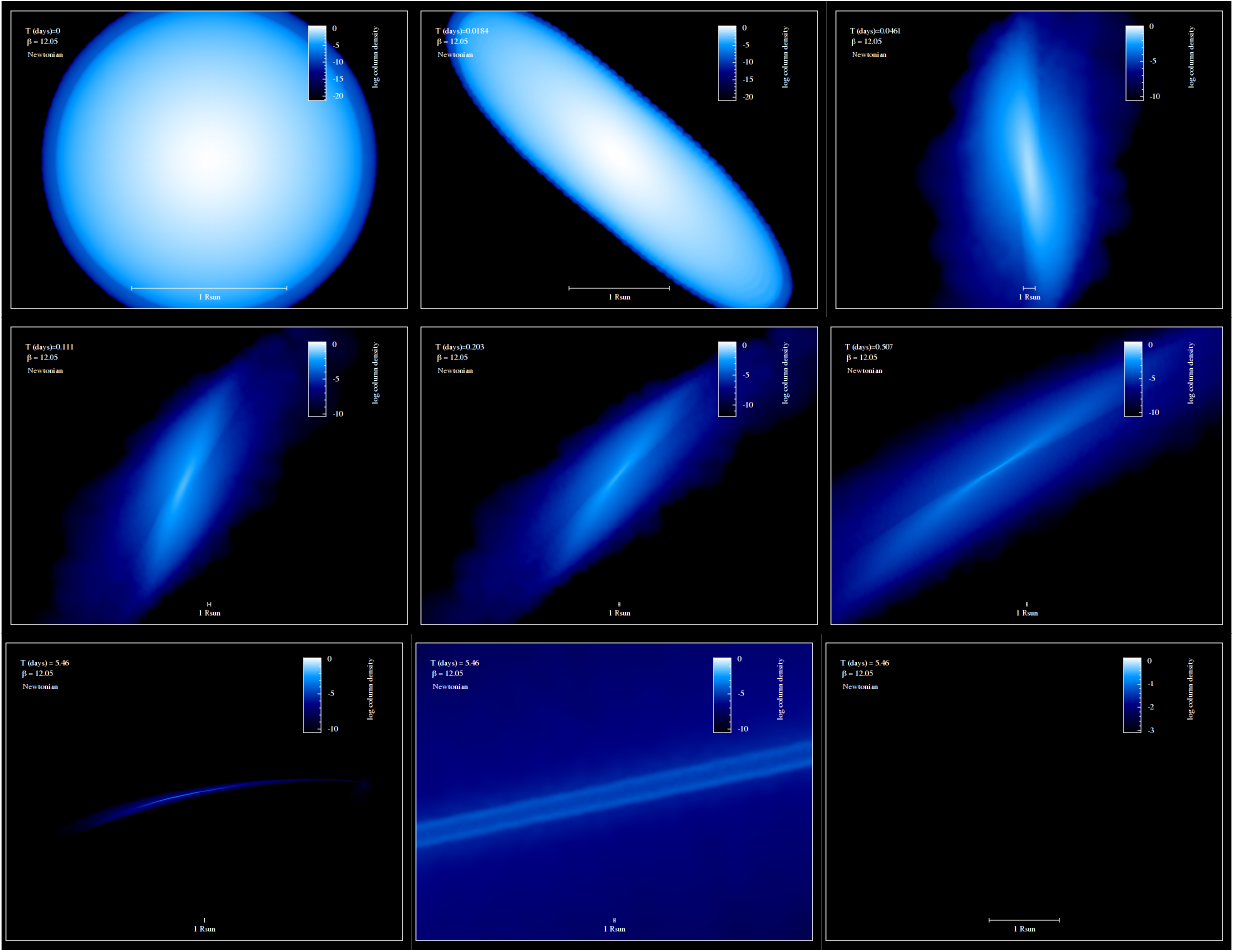}}
\caption
   {
Same as in Fig.~(\ref{fig.SPH_mosaic_beta_2p26_N}) but for a value of $\beta=12.05$
and different times of the total integration.
The last panel again shows nothing because the scale of the
column density of the star is set to the same values of Fig.~(\ref{fig.SPH_mosaic_beta_12p05_R}),
which we have chosen to show the survival core of the star. In this Newtonian case,
however, the gas density is so low that nothing appears.
   }
\label{fig.SPH_mosaic_beta_12p05_N}
\end{figure*}

\begin{figure*}
\resizebox{\hsize}{!}
          {\includegraphics[scale=1,clip]{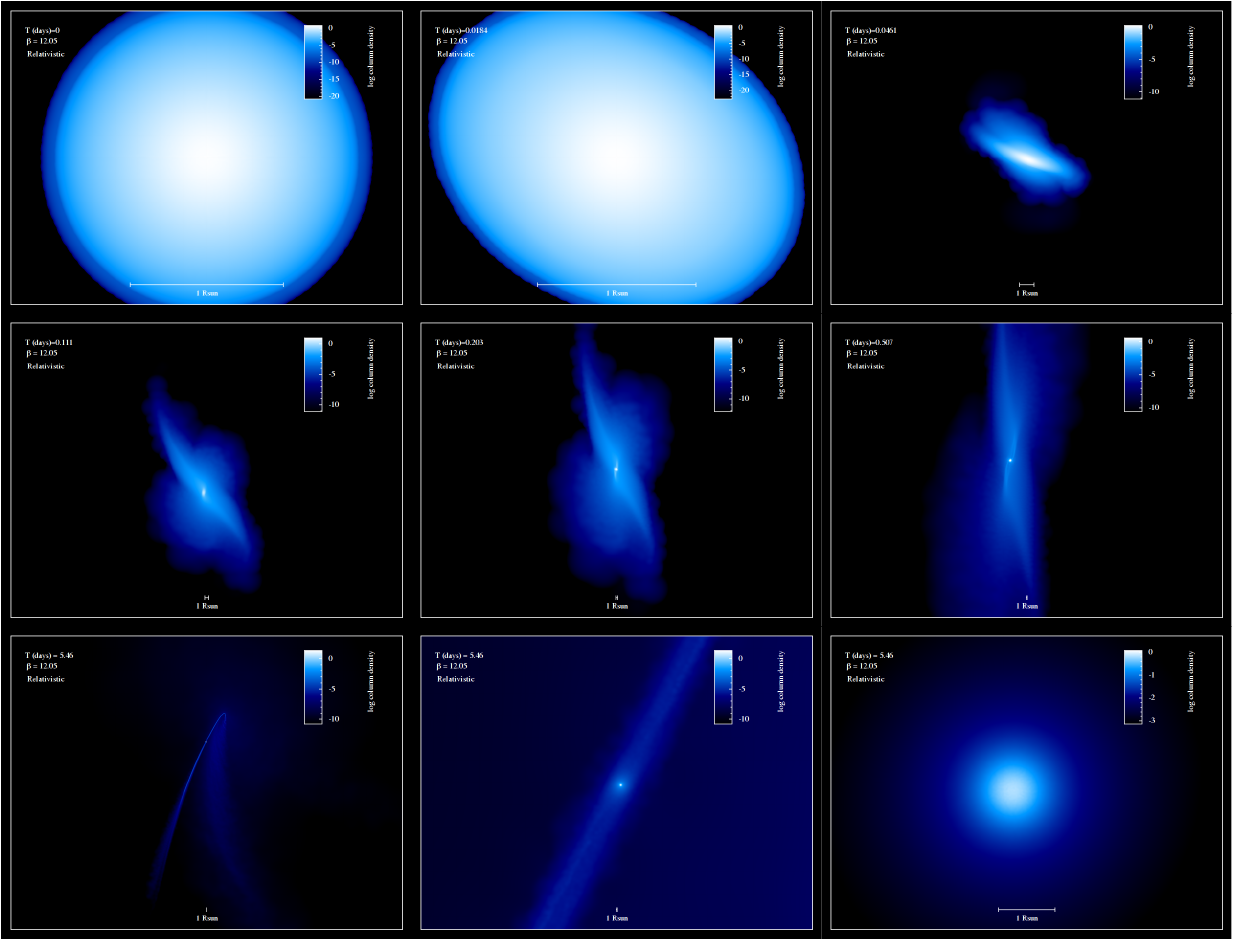}}
\caption
   {
Same as Fig.~(\ref{fig.SPH_mosaic_beta_12p05_N}) for the relativistic counterpart. Already
at early times, the star in the relativistic cases suffers a less significant
spread in size. Also in this case, a survival core is found clearly in the simulation.
   }
\label{fig.SPH_mosaic_beta_12p05_R}
\end{figure*}

This difference becomes even more evident when comparing the extreme case of
$\beta=12.05$, in Fig.~(\ref{fig.SPH_mosaic_beta_12p05_N}), the Newtonian case
and Fig.~(\ref{fig.SPH_mosaic_beta_12p05_R}).  Short after the passage through
periapsis, nothing is left from the original star in the Newtonian case, while
in the relativistic one we find a surviving core at much later times.

\section{Opening of the debris}

\begin{figure*}
          {\includegraphics[width=1.0\textwidth,center]{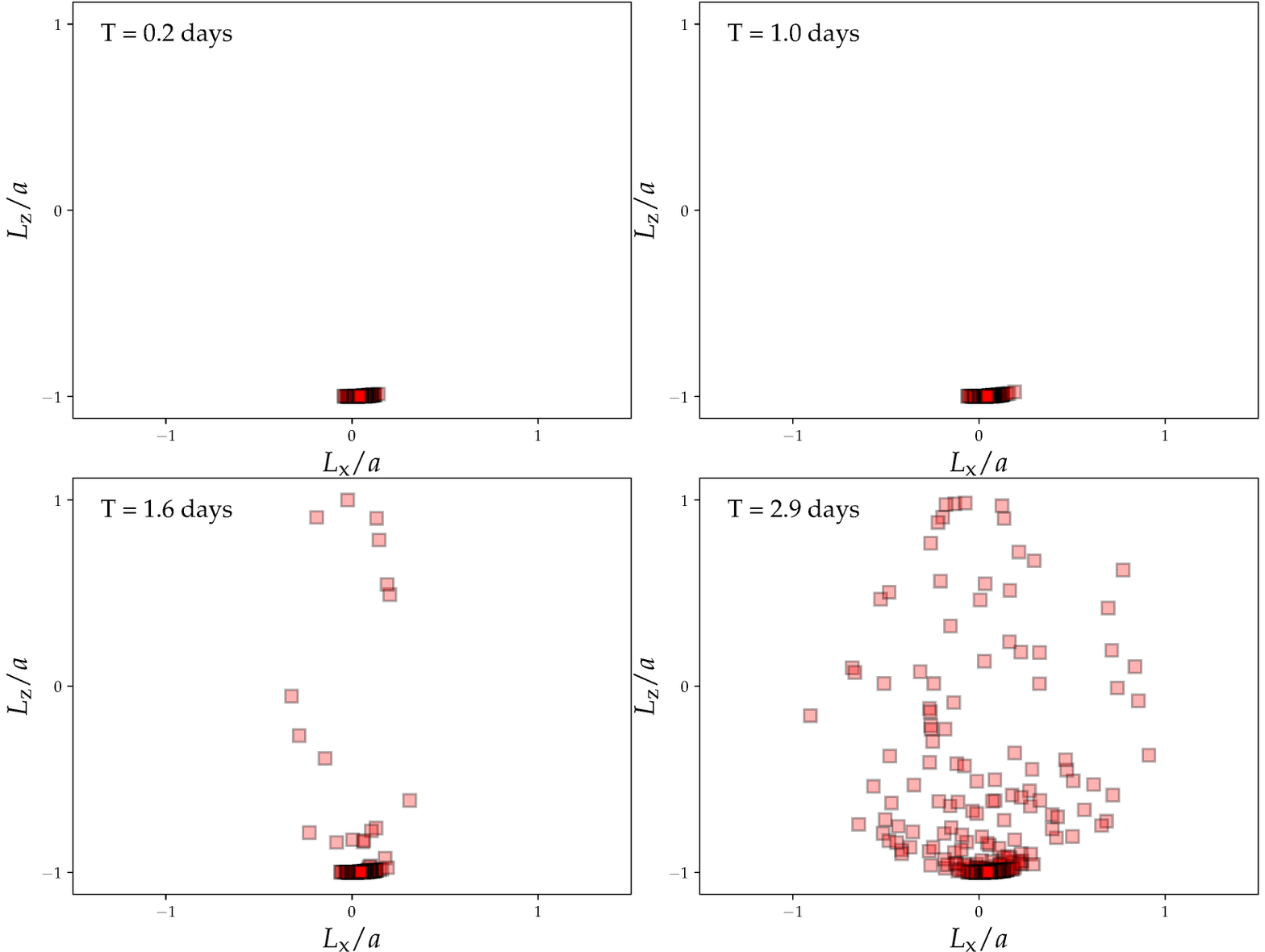}}
\caption
   {
Components z and x of the angular momentum of the gas particles normalised to the spin
of the MBH after the star has passed through periapsis. From the top to the right, we
can see the evolution of the debris as a function of time in days.
   }
\label{fig.Lz_Lx_beta4}
\end{figure*}

In the simulations with a spin component, the debris opens and expands outside
of the initial plan of the orbit. In Fig.~(\ref{fig.Lz_Lx_beta4}) we display
this feature for one of the relativistic cases with spin. We show how the
z-component of the angular momentum $L_z$ normalised to the spin of the MBH
evolves in time as compared to the x component, $L_x$, and we note that the
results are similar for the y component. This outcome is particularly
interesting because it potentially can lead to the intersection of the gas as
it falls back on to the MBH with parts of the same debris which are closer to
the MBH. Such shocks can lead to afterglow flares which are potentially
observable.

However, one cannot solve this problem with an SPH-based code, since we cannot
solve shocks. On the other hand, the amount of time needed for integration for
the gas particles to achieve the maximum distance and fall back on to the MBH
is too long. The accumulation of numerical error and required integration time
makes the exercise pointless. This is indeed why we have integrated
analytically the evolution of the gas particles in the first place, as we
explained before.  One could come up with the idea of converting these
particles again into blobs of gas after the analytical integration but then the
question remains open as to what thermodynamics those clumps of gas should
follow. It would be wrong to assume that their thermodynamical equation of
state is the same as it was when the star underwent the tidal disruption. For
all of these reasons, we just indicate here that the gas does spread out after
the disruption, but we do not make any attempt at trying to continue the
simulation to follow any potential interesting electromagnetic emission.

\section{Properties of the surviving core}

\begin{figure*}
          {\includegraphics[width=1.0\textwidth,center]{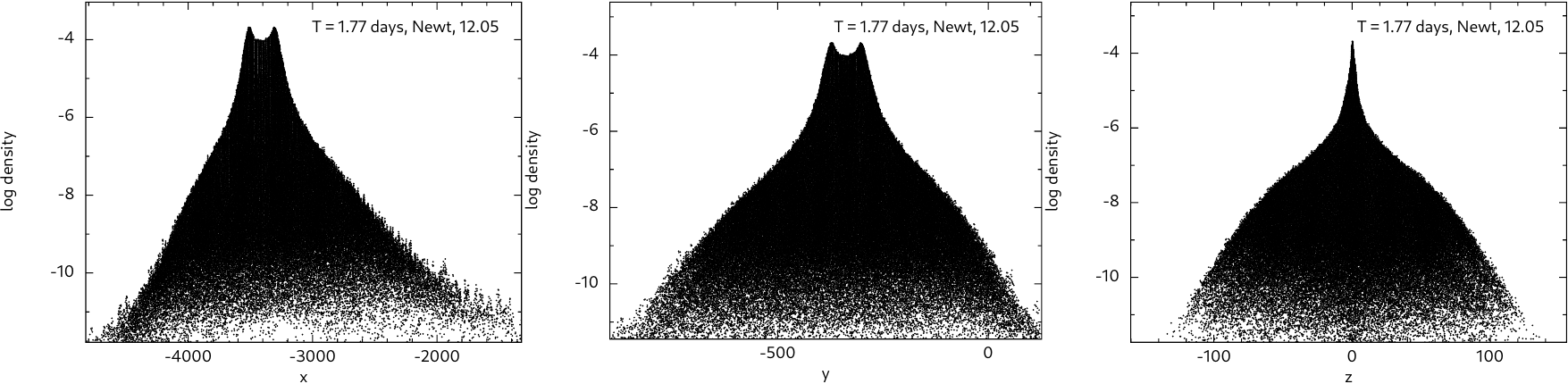}}
\caption
   {
Projection of the density of the gaseous debris in the X-, Y- and Z axis for $\beta=12.05$ in 
the Newtonian case after $1.77$ days. 
   }
\label{fig.Density_Beta_12_Newt}
\end{figure*}

\begin{figure*}
          {\includegraphics[width=1.0\textwidth,center]{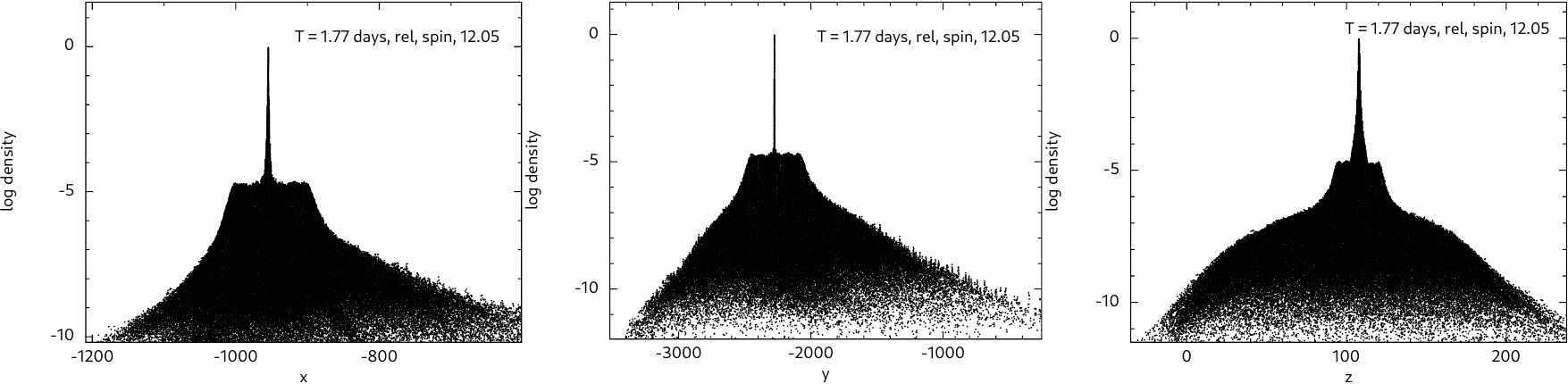}}
\caption
   {
Same as Fig.~(\ref{fig.Density_Beta_12_Newt}) bur for the relativistic case. The
spike corresponds to the surviving core.
   }
\label{fig.Density_Beta_12_RelSpin}
\end{figure*}

In this section we evaluate the properties of the core in the Newtonian and
relativistic case for the most extreme case we have studied, $\beta=12.05$. In
Figs.~(\ref{fig.Density_Beta_12_Newt}, \ref{fig.Density_Beta_12_RelSpin}) we
show the projection in the X- and Y-plane of the column density for both the
Newtonian and relativistic cases at approximately the same time. As in the
previous sections, we can observe a peak of density in the relativistic case,
pinpointing the location of the surviving core. It is also interesting to observe that
the relativistic case has a more complex structure than the Newtonian one, due
to the twisting of the spin acting on the internal structure of the star.

\subsection{An extreme case}
\label{sec.extreme}

\begin{figure*}
          {\includegraphics[width=1.0\textwidth,center]{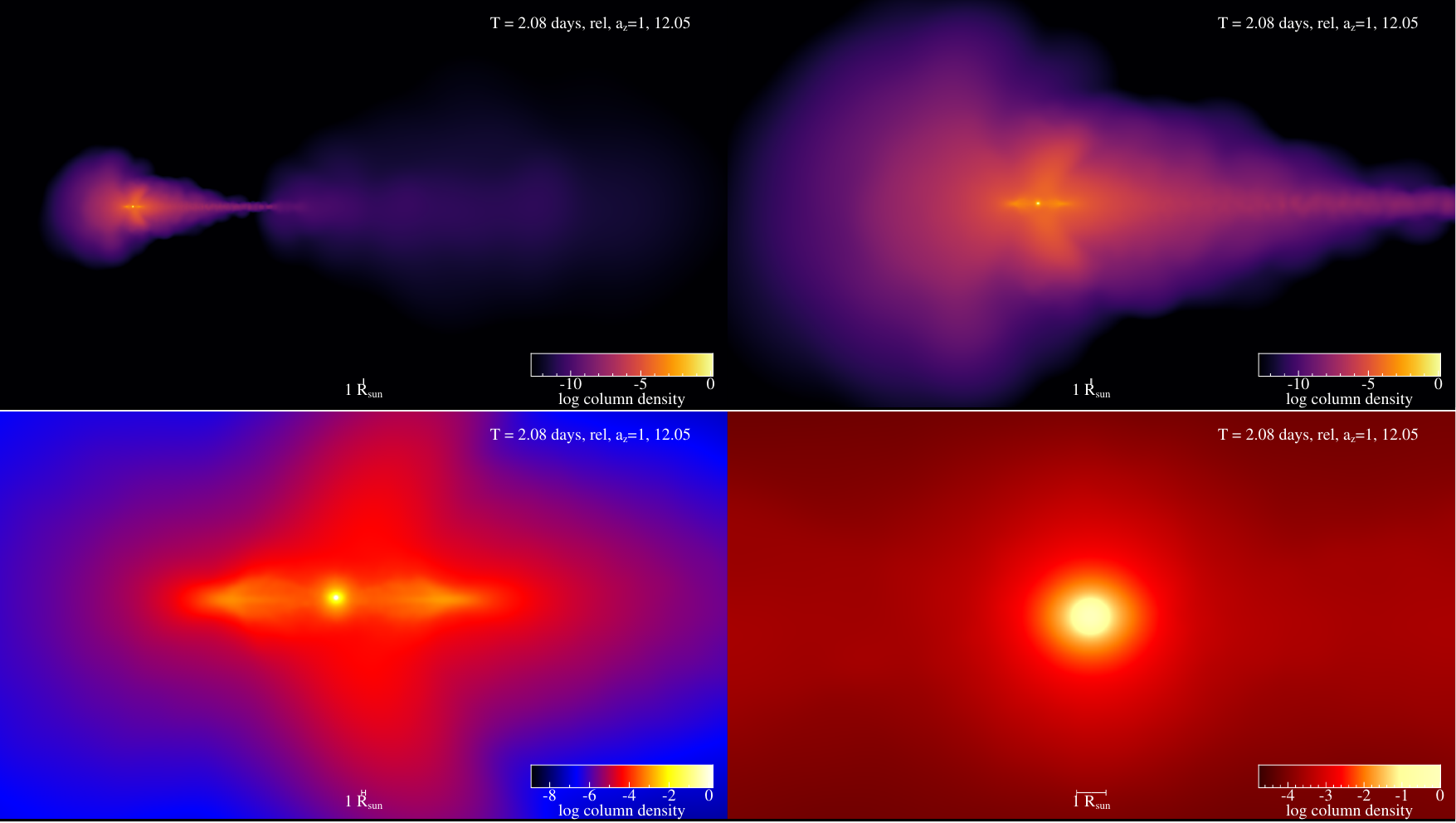}}
\caption
   {
Density of the gaseous debris and surviving stellar core for the case in which
the massive black hole is maximally spinning around the z-axis, in the relativistic
case, after $2.08$ days, for $\beta=12.05$. From the top to the right, we show a 
progressive zoom-in of the projection to visualize the surviving core which, as in
the previous cases, has a size of about half a solar radius. Note that the column
density ranges from different values as we zoom in, so as to define a more clear
depiction of the gaseous particles in each panel.
   }
\label{fig.SpinZ_1}
\end{figure*}

In order to understand how much the orientation and magnitude of the spin affect
the general results, we address the evolution of a star which is tidally disrupted in the deepest
penetration factor, $\beta=12.05$, that we have considered so far for a
particular configuration, in which the massive black hole has a spin value of
$1$ in the z-direction. In Fig.~(\ref{fig.SpinZ_1}) we zoom in from the largest
possible picture to the size of the core. As in all previous relativistic cases,
we see that the general results remain the same. 

Although the structure of the star may appear to be, and is, in fact, different, it is dangerous to try to over-interpret the consequences and implications of this fact. We should remember that we are using an SPH method which, by construction, cannot or does not properly resolve shocks in the system under study. On the other hand, even if we were to study the evolution of the system over very long times to see if the debris intersects, it would be difficult to understand how to translate that fact into observables, since we would have accumulated a non-negligible numerical error. The alternative of analytically integrating the ballistic trajectories of the gas particles to later re-form hydrodynamic structures is unrealistic, as mentioned before, since we would have no information on the thermodynamic properties that these structures should have once the gas particles agglomerate.

\subsection{Fate of the core}

Indeed, for the above reasons, in Sec.~(\ref{sec.extreme}), although it is interesting to understand the subsequent evolution of the core, in particular to see if it remains bound to the MBH, so that it returns to the pericentre of the orbit to undergo further disruptions, it does not make sense to do so numerically. Therefore, we integrate the system until a reasonable time, i.e., until the core of the star is at such a distance from the MBH that any relativistic effect is clearly negligible. Typically, this distance is at least a thousand solar radii. At this point we calculate the centre of mass of the core, to determine its coordinates and velocities, as well as its mass. Assuming that the core follows a Keplerian orbit, we integrate its evolution to determine how long it will take to reach both the apocentre of the orbit and the pericentre. 
In the case that the core is bound and returns to the MBH, we would have to investigate the role played by the dissipative terms of the post-Newtonian approximation. However,
in all the cases we have presented in this paper, the core resulting from the TDE is not bound to the MBH, so that the possibility of successive TDEs is ruled out. How this depends on the initial conditions and the limitations of the numerical method we have used needs to be investigated in detail. However, all this would result in much more dense work than what we are already presenting and will be investigated elsewhere.

\subsection{The role of the number of particles in the hydrodynamical simulations}
\label{sec.resolution}

\begin{figure*}
          {\includegraphics[width=1.0\textwidth,center]{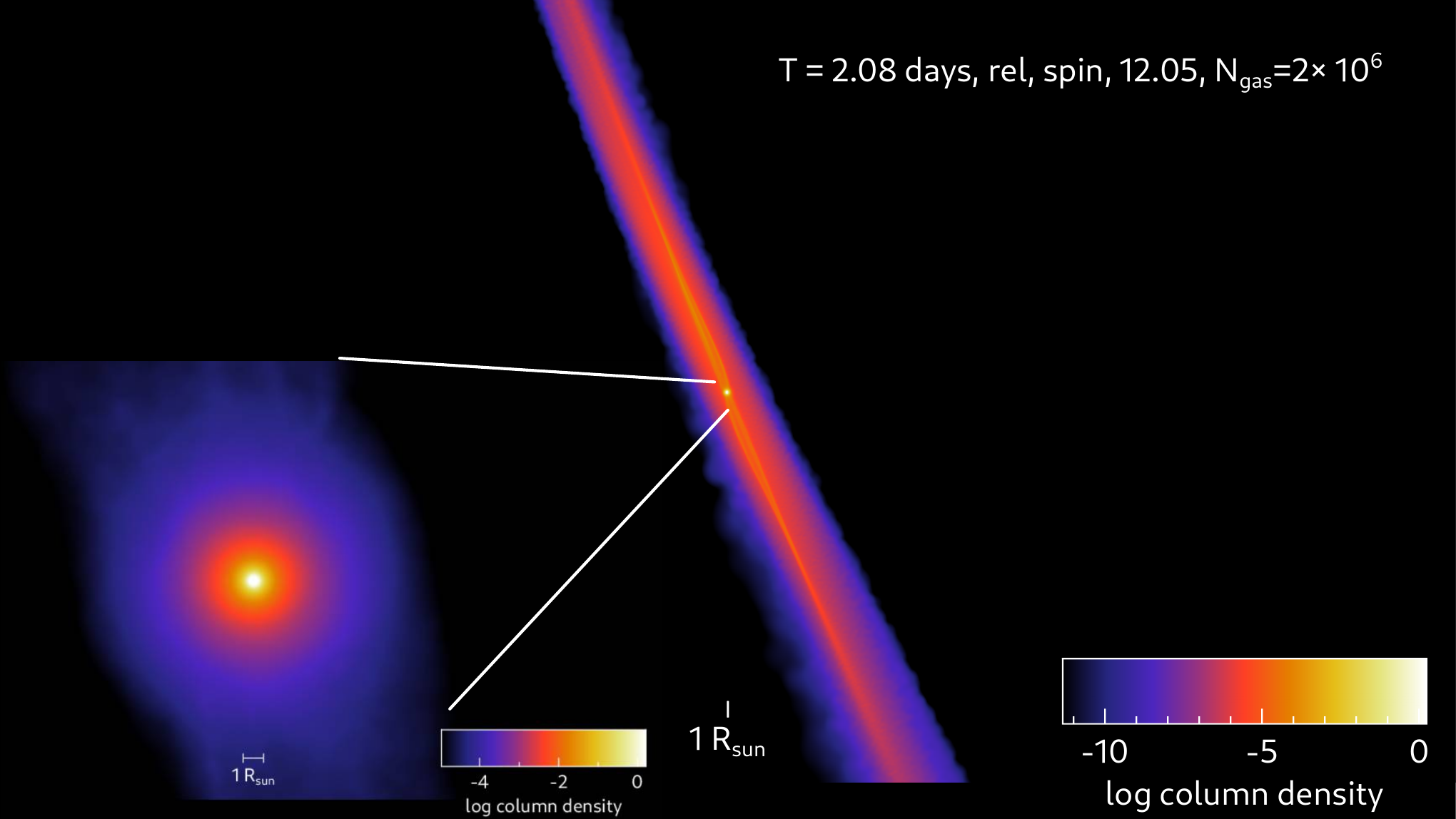}}
\caption
   {
Same as Fig.~(\ref{fig.SPH_mosaic_beta_12p05_R}) but for $2\times 10^6$ gas particles.
   }
\label{fig.2e6particles_beta12_rel}
\end{figure*}

\begin{figure*}
          {\includegraphics[width=1.0\textwidth,center]{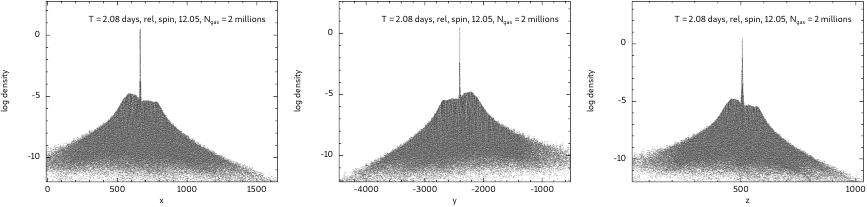}}
\caption
   {
Same as Fig.~(\ref{fig.Density_Beta_12_RelSpin}) but for $2\times 10^6$ gas particles.
   }
\label{fig.Density_Beta_12_RelSpin_2e6_part}
\end{figure*}

So as to see whether the number of particles we have used previously is enough
to capture the physics of the problem, we have performed an additional
simulation with a total of $2\times 10^6$ particles. Going beyond this number
led to memory problems in the computer that we used to create the initial
conditions. In Fig.~(\ref{fig.2e6particles_beta12_rel}) we show the usual
density projection of the gas debris and surviving stellar core. In
Fig.~(\ref{fig.Density_Beta_12_RelSpin_2e6_part}) we depict the density as a
function of the three different axes. When comparing these results to their
lower-resolution counterparts, i.e. Fig.~(\ref{fig.SPH_mosaic_beta_12p05_R})
and Fig.~(\ref{fig.Density_Beta_12_RelSpin}), we can see that the results we
derived previously hold.

\section{Distance between geodesics of particles on a parabolic orbit around a spinning massive black hole}

\subsection{Numerical study}

One better way to try to understand why a core survives in the relativistic
simulations is to calculate the geodesic convergence (or divergence) of the
elements of the star as it approaches the pericentre of the MBH.  The idea is
to investigate if in the relativistic case the particles, which are to be
envisaged as representative parts of the star, will come closer or not in the
relativistic case as compared to the Newtonian one. In order to do so, we
simulate the trajectory of six test particles using the numerical programme
ARChain \citep{MM06,MikkolaMerritt2008}.  This programme does not take into
account hydrodynamics, but pure, point-like dynamics. The advantage is that it
is very accurate and features the implementation of the post-Newtonian terms as
first presented in \cite{KupiEtAl06}, which is the same we have used in the
hydrodynamical experiments.

\begin{figure*}
          {\includegraphics[width=1.0\textwidth,center]{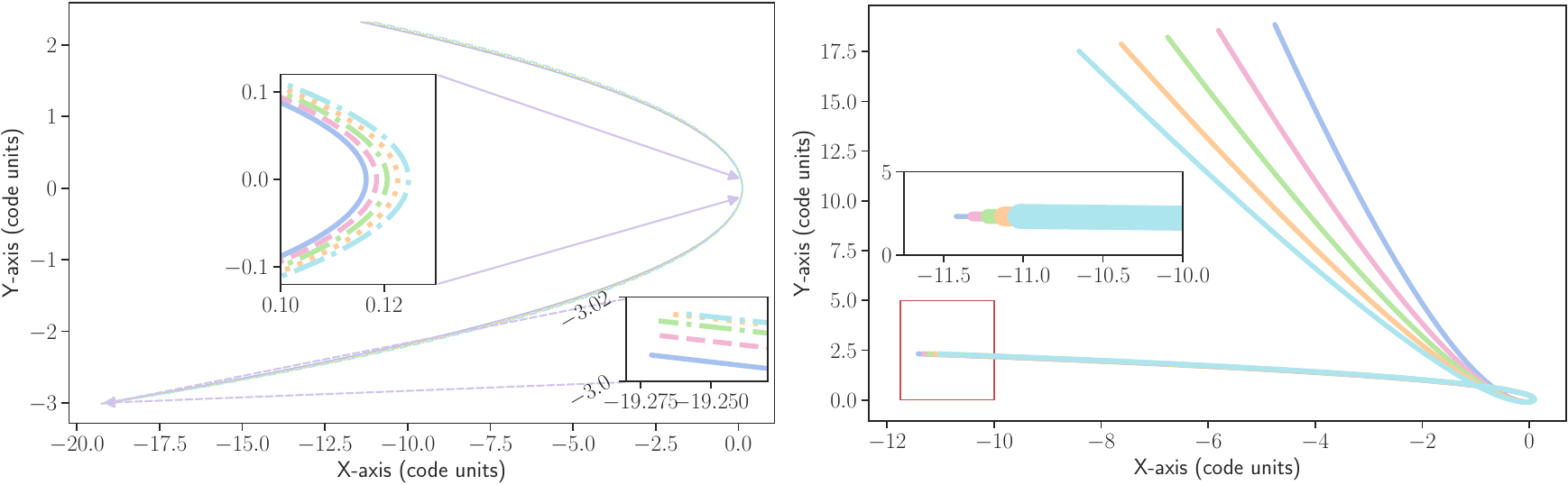}}
\caption
   {
Trajectories in the XY plane of six test particles around a MBH in the
Newtonian- (left) and relativistic (right) cases. In both panels we display the
evolution of the particles and a zoom-in at the pericentre for the Newtonian
case. The relativistic one also includes a zoom-in of the initial trajectories
of the particles, which are slightly shifted but are otherwise identical.  The
MBH is located initially at the origin.  The test particles have a mass
$10^{-14}$ times smaller than than of the MBH.  The initial penetration factor
is set to $\beta=1.2$ for numerical reasons, since the programme will declare
the particles as relativistic mergers for deeper factors.
   }
\label{fig.ARChain_Six_Components}
\end{figure*}

In Fig.~(\ref{fig.ARChain_Six_Components}) we show the Newtonian and
relativistic case without spin for six test particles on an initially parabolic
orbit around a MBH.  I.e. we want to see whether the particles tend to get
closer in the relativistic case of this ``dust star'' as compared to the
Newtonian case. As we can see in the figure, in the Newtonian case the
particles follow the usual Keplerian trajectory (modulo some fluctuations due
to their mutual attraction). The relativistic case undergoes precession and the
particles follow ballistic trajectories after the periapsis passage. That
ulterior evolution is not representative of what different parts in a star
would do, since, as explained, we are not taking into account hydrodynamics.
The timescale for the star to readjust to a perturbation caused by the
gravitational interaction with the black hole is determined by the star's
sound-crossing time, which is the time it takes for a sound wave to travel
across the star's diameter. For a typical main-sequence star with a radius of
$R_\odot \approx 6.96\times10^8$ m and a sound speed of $c_s \approx 10^5$ m/s,
the sound-crossing time is on the order of a few minutes. In the dynamical
simulations, the time for a particle to go though the periapsis distance is
smaller.

It is important to note that for this comparison to make sense, we need to
focus at what happens at periapsis, and not much farther beyond that point.
However, the timescale wich dominates at periapsis is the relativistic one, so
that we can neglect hydrodynamics and assume point particles interacting only
via gravity.

\begin{figure*}
          {\includegraphics[width=1.0\textwidth,center]{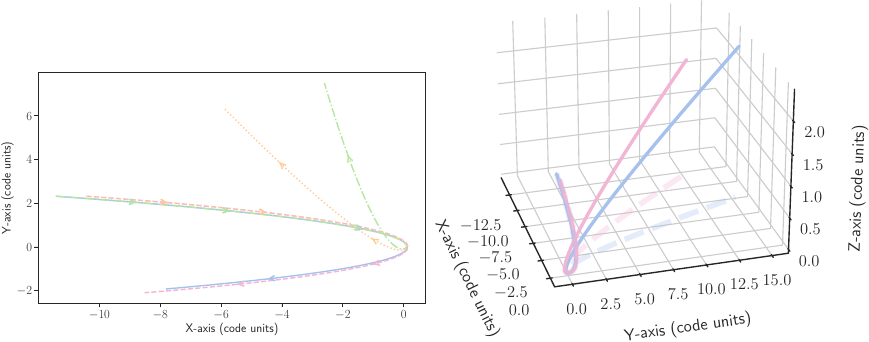}}
\caption
   {
Trajectories in a volume XYZ of two test particles around a MBH for the
relativistic case without (left) and with spin (right). In the second case, the
trajectories leave the initial plane of the orbit and obtain a z component. We
display in arrows the sense of the motion of the test particles in the upper,
left panel.
   }
\label{fig.ARChain_Two_Components_Trajectories}
\end{figure*}

\begin{figure*}
          {\includegraphics[width=1.0\textwidth,center]{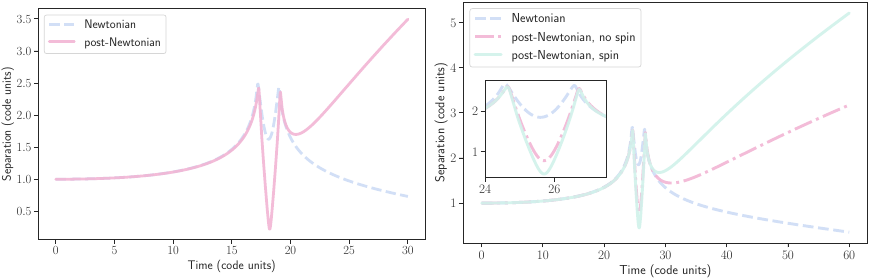}}
\caption
   {
Difference between two particles along their orbit around a MBH in the
Newtonian and relativistic case. In the latter,
we also display the case with and without spin for the MBH (right panel). We include
a zoom-in in the right panel to clearly show the difference at periapsis
passage.
   }
\label{fig.ARChain_Difference_Two_Components}
\end{figure*}

In Fig.~(\ref{fig.ARChain_Difference_Two_Components}) we depict the difference
between the orbits in a two-particles case. We include two different
relativistic treatments; one which does not take into account the spin of the
MBH and another which does. As we can see, in the second case, the difference
is the largest, meaning that the particles are getting a factor two of
compression in the relativistic case when compared to the Newtonian one. Beyond
the periapsis, the three different cases have very different evolutions
because, as previously explained, we are in the purely dynamical regime; i.e.
we do not take into account the hydrodynamics. It is interesting to see how the
peaks in the difference of the relativistic cases are shifted as compared in the
Newtonian one, since a smaller pericentre distance leads to a larger velocity and
hence a shorter amount of time spent in that region.

The purpose of these dynamical test particle simulations is to illustrate why,
in the relativistic case, particularly if we have spin, the star has a
surviving core. As the star passes through periapsis, it will undergo an
internal compression that will increase the density of the star, particularly
at its centre. This result does not prove, but suggests, that relativistic
effects could indeed compress the star enough to allow some of it to survive
the tidal stresses as a self-gravitating object. This is what we observe in
Figs.~(\ref{fig.Bound_Ejected_Mass_Beta1p64_to_3p62}) and
(\ref{fig.Bound_Ejected_Mass_Beta5p15_and_12p05}).

\subsection{Analytical study}

Now that we have seen that the dynamical study of the dust star leads to
smaller distances at periapsis, confirming the results of the hydrodynamical simulations, what remains is to understand why this is the case; i.e. what is the physical reason for it. For this, let us consider the equation of geodesic deviation in proper time, 

\begin{align}
\frac{D^2 \xi^\text{r}}{d\tau^2} = \Big[ &
                              R^\text{r}_\text{ttr} \left( \frac{dt}{d\tau} \right)^2
                            + R^\text{r}_{\theta \theta r} \left( \frac{d\theta}{d\tau} \right)^2 \nonumber \\
                            & + R^\text{r}_{\phi \phi r} \left( \frac{d\phi}{d\tau} \right)^2
                            \Big]\,\xi^\text{r},
\end{align}

\noindent
where

\begin{equation}
\frac{D}{d\tau} = \nabla_{\mathbf{u}} = u^{\mu} \nabla_{\mu}
\end{equation}

\noindent
is the covariant derivative along the observer's four-velocity;
and

\begin{align}
R^\text{r}_\text{ttr}        & = \frac{2G \, m_{\bullet}}{r^3} \left( 1 - \frac{2G \, m_{\bullet}}{r c^2} \right) \nonumber \\
R^\text{r}_{\theta \theta r} & = \frac{G m_{\bullet}}{r c^2}\\ \nonumber
R^\text{r}_{\phi \phi r}     & = \frac{G m_{\bullet}}{r c^2} \sin^2{\theta},
\label{eq.ricci}
\end{align}

\noindent 
see e.g. \cite{Chandrasekhar1998}, and we have replaced the Schwarzschild radius with

\begin{equation}
R_{\rm Schw} = 2 \frac{G \,m_{\bullet}}{c^2}.
\end{equation}

\noindent 
It is important to note that, since we are neglecting the role of hydrodynamics
and self-gravity, the test particles which conform the dust star strictly
follow geodesics. We consider a co-moving observer along these geodesics (i.e., an observer in the Fermi or free-fall coordinates), located on one of the test particles who observes a locally flat spacetime. Thus,

\begin{align}
&\frac{dt}{d \tau} = c^2 \\
&\frac{dx^i}{d \tau} = 0 \;\;\ \text{for any spatial coordinate},
\end{align}

\noindent
and hence

\begin{align}
    \dfrac{D^2}{d \tau^2} &= \frac{d^2}{dt^2} + \frac{2G \, m_{\bullet}}{r^2} \left( 1 - \frac{2G \, m_{\bullet}}{c^2 r^2}  \right)^{-1} \frac{d}{dt} + \nonumber \\
    & \left( \frac{2G \, m_{\bullet}}{r^2} \right)^2. 
\end{align}

As a result, the geodesic deviation equation becomes

\begin{align}
&\frac{d^2 \xi^\text{r}}{dt^2} + \frac{2G \, m_{\bullet}}{r^2} \left( 1 - \frac{2G \, m_{\bullet}}{r c^2}  \right)^{-1} \frac{d \xi^\text{r}}{dt} \nonumber \\
&+\Big[ \left( \frac{2G \, m_{\bullet}}{r^2 c^2} \right)^2 - \frac{2G \, m_{\bullet}}{r^3} \left( 1 - \frac{2G \, m_{\bullet}}{r c^2} \right) \Big] \xi^\text{r} = 0,
\end{align} \label{eq:geodev}

\noindent 
which is a second-order linear, homogeneous differential equation. Such an equation admits the general solution of the following type

\begin{equation}
\xi^r = C_1 e^{\lambda_1\,t} + C_2 e^{\lambda_2\,t},
\end{equation}

\noindent 
with $C_1$ and $C_2$ two constants which are a result of the initial conditions and $\lambda_1$ and $\lambda_2$ two exponents that must fulfil

\begin{align}
\lambda^2 &+ \frac{2G \, m_{\bullet}}{r^2} \left( 1 - \frac{2G \, m_{\bullet}}{r c^2}  \right)^{-1} \lambda \nonumber \\
&+\Big[ \left( \frac{2G \, m_{\bullet}}{r^2 c^2} \right)^2 - \frac{2G \, m_{\bullet}}{r^3} \left( 1 - \frac{2G \, m_{\bullet}}{r c^2} \right) \Big] = 0.
\end{align}

\noindent
This leads to the solutions

\begin{align}
\lambda_1 & = -\frac{G m_{\bullet}}{r^2 c^2} \left( 1 - \frac{2 G m_{\bullet}}{r c^2} \right)^{-1} + \nonumber \\
& \left[ \frac{G m_{\bullet}}{c^2 r^4} \left(\frac{G m_{\bullet}}{c^2} \left[\frac{r}{\left(1 - \frac{2 G m_{\bullet}}{r c^2}\right)^2} - 20\right] + 8 r \right) \right]^{1/2} \\
\lambda_2 & = -\frac{G m_{\bullet}}{r^2 c^2} \left( 1 - \frac{2 G m_{\bullet}}{r c^2} \right)^{-1} - \nonumber \\
& \left[ \frac{G m_{\bullet}}{c^2 r^4} \left(\frac{G m_{\bullet}}{c^2} \left[\frac{r}{\left(1 - \frac{2 G m_{\bullet}}{r c^2}\right)^2} - 20\right] + 8 r \right) \right]^{1/2},
\end{align}

\noindent 
which are always real when $r > R_{\rm Schw}$.

We can derive the values of the two constants considering the initial conditions. Since the two test particles are part of the dust star, initially the distance
between the two is a fixed quantity which, at most, admits a value of $2R_{\odot}$ since
that is the diameter of the star; also, at time $t=0$, the change in the distance is zero.
I.e. the particles are neither moving away from each other nor closer because the star
is considered to be in equilibrium. Therefore, 

\begin{align}
\xi^r(0) & \leq 2R_{\odot}, \nonumber \\
\frac{d\xi^r}{dt}\Bigg|_{(0)} & = 0.
\end{align}

\noindent 
We therefore have that

\begin{align}
\xi^r(0) &= C_1 + C_2 \leq 2R_{\odot}, \\
\frac{d\xi^r}{dt}\Bigg|_{(0)} & = \lambda_1\,C_1 + \lambda_2\,C_2 = 0.
\end{align}

\noindent 
From the last equation we obtain

\begin{align}
C_1 &= \frac{\xi^r(0)}{2} \left( 1 - \chi \right) \\
C_2 &= \frac{\xi^r(0)}{2} \left( 1 + \chi \right),
\end{align}

\noindent
where we have defined

\begin{align}
    \chi := \frac{r^2 G m_{\bullet}}{2} & \Big[ G m_{\bullet} \left(2 c^6 r^3+40 c^2 G^2 m_{\bullet}^2 r  + \right. \nonumber \\
    &\left. \left(c^2-16\right) c^4 G m_{\bullet} r^2 - 32 G^3 m_{\bullet}^3\right) \Big]^{-1/2}.
\end{align}

\noindent
This quantity drops rapidly as $r$ grows. This means that, far away from the MBH, the two coefficients tend to the same. As a result, both exponential solutions ($e^{\lambda_1 t}$ and $e^{\lambda_2 t}$) are non-zero and neither can be assumed to be very small and negligible. 

\begin{figure}
\resizebox{\hsize}{!}
          {\includegraphics[scale=0.2,clip]{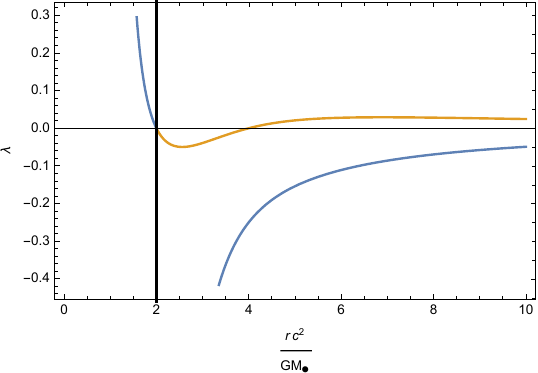}}
\caption
   {
Evolution of the two exponents resulting from the solution of the geodesic deviation
equation, $\lambda_1$ (blue curve) and $\lambda_2$ (orange curve), with respect to the radial distance (normalised by half $R_{\rm Schw}$). The MBH Schwarzschild radius is noted with a black vertical line.
Within this radius both solutions are identical and represented with the
same blue curve. We name the region corresponding to the range of values for which both exponents are negative the maximum compression zone.
   }
\label{fig.exponents}
\end{figure}

At this point, in order to examine which of the exponents is more significant, it is useful to examine their behaviour with respect to the radial distance. Notably, one solution ($\lambda_1$) is always negative for $r > R_{\rm Schw}$; hence the corresponding exponential ($e^{\lambda_1 t}$) refers to a shrinking solution for any radial distance of the star from the MBH. The second solution ($\lambda_2$), however, alternates sign; it is negative when $R_{\rm Schw} < r < 2 R_{\rm Schw}$, and positive when $r > 2 R_{\rm Schw}$. As a result, the corresponding exponential solution is shrinking when the particles (i.e., the star) are close to the MBH, and expanding when they are way from it, as we can see in Fig.~\ref{fig.exponents}.

As we can see from Fig.~\ref{fig.exponents}, the effect of convergence
of geodesics is maximised when $R_{\rm Schw} < r < 2 R_{\rm Schw}$,
which we dub the maximum compression zone,
because both exponents are negative. However, not all test particles will necessarily fall this close to the MBH. The star will generally have a larger penetration factor. For larger radii, one of the
two exponents is not negative anymore (the orange curve, $\lambda_2$).
In order to understand how the distance between the two test particles will evolve in the case in which we are outside of the maximum compression zone, we observe that in Eq.~(\ref{eq:geodev}) the coefficient of ${d \xi^r}/{dt}$ is larger for small $r$, but drops faster than the coefficient of $\xi^r$, so that it can be safely ignored outside the maximum compression zone. In this regime, therefore, the geodesic deviation equation becomes

\begin{equation}
    \frac{d^2 \tilde{\xi}^r}{dt^2} + \Big[ \left( \frac{2G \, m_{\bullet}}{r^2 c^2} \right)^2 - \frac{2G \, m_{\bullet}}{r^3} \left( 1 - \frac{2G \, m_{\bullet}}{r c^2} \right) \Big] \xi^\text{r} = 0.
\end{equation}

\noindent
The solution is again of the form

\begin{equation}
    \tilde{\xi}^r = C_1 e^{\lambda_1 t} + C_2 e^{\lambda_2 t},
\end{equation}

\noindent
where the exponents can be determined by the characteristic
equation to be

\begin{align}
    \lambda_1 &= \frac{1}{r^2} \sqrt{2G \, m_{\bullet} r \left( 1 - \frac{4 G \, m_{\bullet}}{r c^2} \right)} \nonumber \\
    \lambda_2 &= -\frac{1}{r^2} \sqrt{2G \, m_{\bullet} r \left( 1 - \frac{4 G \, m_{\bullet}}{r c^2} \right)},
\end{align}

\noindent
and

\begin{equation}
    C_1 = C_2 = \dfrac{\xi^r (0)}{2}
\end{equation}

\noindent
are the constants (determined by the same initial conditions, as before).
These solutions contains two exponential functions; one shrinking and the other one expanding at the same pace. Obviously, for large times (i.e. large $r$), the growing exponential will dominate and the geodesics will deviate, as we saw in the numerical calculations, e.g. Fig.(\ref{fig.ARChain_Two_Components_Trajectories}). This
was indeed expected since we are not taking into account the hydrodynamics of the problem because we do not need it for the short
timescales of interest. Therefore this solution, although mathematically
correct, is physically irrelevant.

In more interesting (shorter) timescales, as the star (and hence the dust particles) is getting closer to the periapsis distance of the MBH, both terms will contribute. One exponent will tend to bring closer together the test particles and the other one to increase their distance. To ponder how these two concurring exponents affect the global evolution of the two geodesics, we introduce a ``half-life'' timescale, a characteristic timescale for an exponent to dominate over the other,

\begin{equation}
    t_{\text{HL}} = \frac{1}{\lambda_1} = \frac{1}{| \lambda_2 |}.
\end{equation}

\noindent
This timescale is calculated to be

\begin{equation}
    t_{\text{HL}} = r^2 \left[ 2 G \, m_{\bullet} r \left( 1 - \frac{4 G \, m_{\bullet}}r c^2 \right) \right]^{-1/2} \propto r^{3/2},
\end{equation}

\noindent
which means that the timescale of the growing exponential to dominate the shrinking one grows with $r^{3/2}$. As a result, the solutions can be expressed as

\begin{equation}
    \tilde{\xi}^r \sim \frac{\xi^r(0)}{2} \left[ \exp\left({\frac{t}{r^{3/2}}}\right) + \exp\left({-\frac{t}{r^{3/2}}}\right) \right].
\end{equation}

\begin{figure}
\resizebox{\hsize}{!}
          {\includegraphics[scale=0.2,clip]{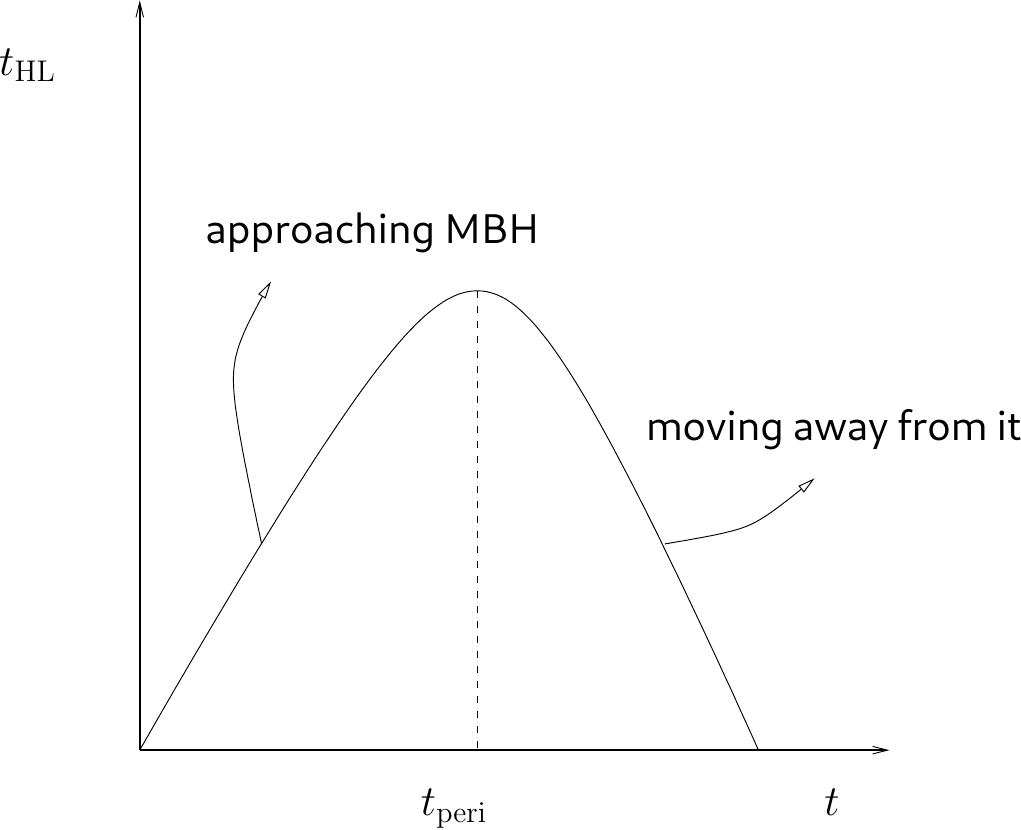}}
\caption
   {
Schematic illustration of the ``half-life'' timescale, which grows as the star approaches the MBH, reaching a
maximum at pericentre. Later it drops as the star moves away from it. During the first half, the geodesics converge, leading to the build up of a central core. The closer the star comes to the MBH the stronger this effect. During the second half, as the star is moving away from the MBH, the geodesics diverge, but this regime is not relevant, as explained in the main body of the text.
   }
\label{fig.thL_xfig}
\end{figure}

\noindent
Of course, as the particles travel in a parabolic orbit, they initially come close to the MBH ($r$ drops over time until they reach periapsis) and then they go away from it ($r$ grows over time after they leave periapsis). 
This effect is captured by the initial growth of the ``half-life'' time, which is followed by a rapid shrinkage, as illustrated in Fig.~\ref{fig.thL_xfig}. 
As a result, as the particles come close to periapsis, the geodesics converge (the closer they come to the MBH, the smaller the distance between them). At later times, and as we discussed previously, the
dust particles will start to deviate in their geodesics, but this
effect is not physical. What matters is the effect on shorter timescales
as the star is approaching periapsis, because the convergence of
geodesics, when generalised to a full star, imply the building up of a core which will withstand the tidal forces of the process and survive
the disruption, as we find in the hydrodynamical simulations.
We note that, although in our configuration the star initially is in the plane of the orbit, in a more general case in which $\theta$ and $\dot{\theta}$ do not vanish, the result remains qualitatively the same,
as we are working from the point of view of a comoving observer.

\section{Conclusions}

In this work we have addressed the problem of TDEs being less luminous than
theoretically expected in the accretion disk model. We run a set of Newtonian
SPH simulations of an unbound star of one solar mass and a MBH of mass
$10^6\,M_{\odot}$ with penetration parameters ranging from $1.64$ to $12.05$.
We re-do the simulations with exactly the same $\beta$ parameters and initial
conditions but taking into account relativistic (post-Newtonian) corrections.
For this we consider two different sets of simulations - One which only
includes the first correction to periapsis shift in the equations of motion and
another which additionally takes into account the spin-orbit coupling
correction up to next-to-lowest order. For $\beta$ values starting at $\beta
\gtrapprox 2.25$, all relativistic simulations feature a surviving core of the
original star. The Newtonian simulations, however, do not. Only the lowest
value of $\beta=1.64$ in the Newtonian case displays a core which does not last
long bound to the original star. As a consequence, the fallback rates are lower
in the corresponding relativistic cases, and hence the luminosity is also
lower.  The deeper the TDE, the bigger the difference in luminosity between the
Newtonian and relativistic simulations. The effect of the spin only plays an
important role, as expected, for extremely deep penetration factors. This was
also noted by the work of \cite{GaftonRosswog2019}, who find that precession leads
to debris configurations which are absent in the Newtonian cases.

Moreover, in the relativistic cases the energy distribution is more spread out,
so that in each specific energy bin there is less matter.  Hence, the fall back
rate in every time step is lower; $dM /dt$ is closely related to $dM/dE$, with
$E$ the specific energy relative to the MBH. This can be seen in e.g. Fig. 3 of
\cite{EvansKochanek1989} and the work of \cite{RyuEtAl2020a,RyuEtAl2020b},
which shows that TDEs in the relativistic case has an energy distribution with
significant wings, as well as Fig. 2 of \cite{RyuEtAl2020c} and
\cite{RyuEtAl2020d} for a full disruption.  If $E$ is wider, $dM/dE$ will be
smaller and, thus, $dM/dt$ as well.

We also study the opening of the debris in the relativistic case and find that the spin
allows it to leave the initial plane of the orbit. The probability of debris from the gas, 
which has been confined within the plane, colliding with itself or with parts of the star 
is high. As the debris follows different ballistic trajectories, the different fractions 
of the star, both on their way to the apocentre and those that have been left behind, 
may eventually interact, producing energetic bursts.

The analysis has been primarily addressed with SPH simulations but complemented
by a dynamical numerical toy model of test particles. For this, we examine the behavior
of geodesics in the context of test particles of a star approaching a black
hole, neglecting the role of hydrodynamics; i.e. a dust star.  We evaluate the distance between
the geodesics of these test particles with a precise dynamical code, ARChain, which
features post-Newtonian corrections as the ones we have implemented in the hydrodynamical
code, since the scheme is based as well in the work of \cite{KupiEtAl06}. This exercise
is useful because we can neglect the role of the hydrodynamics in the regime in which
the relativistic corrections are important, i.e. during the periapsis passage. The
idea is that the timescale in which hydrodynamics plays a role is longer than the timescale
during which relativity plays a role. Hence, the results are to be interpreted only during
the periapsis passage.
The trajectories of the particles represent the tidal deformation experienced by the star as
it approaches the black hole. We find that this distance decreases in the
relativistic case, in particular in the one with spin, as the star gets closer
to the black hole, indicating that the star is being stretched by the tidal
forces. This stretching effect is a key factor in the tidal disruption of the
star. It is important to note that we find that this distance remains finite,
indicating that the star does not get completely disrupted but retains a core.
Finally, we investigate these results analytically using the relativistic equation of geodesic
deviation and confirm the numerical findings; i.e. different parts of the star experience a compression
during periapsis passage which is responsible for the building up of a denser core which
survives the disruption, contrary to the Newtonian calculations.

Our results suggest that in Nature TDEs must have deeper penetration parameters
than previously thought to explain the observations. These orbits naturally
lead to the consequence of a reduced observed luminosity regardless of the
accretion disc, simply due to the fact that relativity allows a part of the
star to survive the disruption. 

\section*{Acknowledgments}

The initial idea of this work was presented as a talk at the Al{\'a}jar meeting
in
2013\footnote{\href{http://astro-gr.org/alajar-meeting-2013-stellar-dynamics-growth-massive-black-holes/}{Workshop Alájar 2013.}}
and later as an invited talk at the workshop ``TDE17: Piercing the sphere of
influence''\footnote{\href{https://www.ast.cam.ac.uk/meetings/2017/tde17.piercing.sphere.influence}{Workshop Cambridge 2017.}}
which took place in Cambridge. PAS thanks Enrico Ram{\'\i}rez-Ruiz, Pablo
Laguna, Zoltan Haiman, Julian Krolik, Nick Stone, Sterl Phinney, Ramesh
Narayan, and Stephan Rosswog for comments during the workshop, as well as
Julian Krolik for pointing him the publications by him and his collaborators.
The simulations performed with ARChain were possible thanks to Seppo Mikkola,
who provided us with a copy of his code.

\section*{Data Availability}    
Any data used in this analysis are available on reasonable request from the first author.

\label{lastpage}
\end{document}